\documentclass[lettersize,journal]{IEEEtran}
\usepackage[utf8]{inputenc}
\usepackage[T1]{fontenc}

\usepackage{enumitem}
\usepackage{amsmath, amssymb}
\usepackage{semantic}
\usepackage{nicematrix}
\usepackage[caption=false,font=normalsize,labelfont=sf,textfont=sf]{subfig}
\usepackage{arydshln}
\usepackage{algpseudocode}
\usepackage{algorithm}
\usepackage{graphicx}
\usepackage{comment}
\usepackage{booktabs}
\usepackage{pifont}
\usepackage{balance}
\usepackage{multirow}
\usepackage{hhline}
\usepackage{optidef}
\usepackage{relsize}
\usepackage{mathrsfs}
\usepackage{tablefootnote}
\usepackage{float}
\usepackage{hyperref}
\usepackage{cite}
\usepackage{makecell}
\usepackage{varwidth}
\usepackage{tikz,xcolor,hyperref}
\definecolor{lime}{HTML}{A6CE39}
\DeclareRobustCommand{\orcidicon}{
	\begin{tikzpicture}
	\draw[lime, fill=lime] (0,0) 
	circle [radius=0.16] 
	node[white] {{\fontfamily{qag}\selectfont \tiny ID}};
	\draw[white, fill=white] (-0.0625,0.095) 
	circle [radius=0.007];
	\end{tikzpicture}
	\hspace{-2mm}
}
\foreach \x in {A, ..., Z}{\expandafter\xdef\csname orcid\x\endcsname{\noexpand\href{https://orcid.org/\csname orcidauthor\x\endcsname}
			{\noexpand\orcidicon}}
}
\usepackage[compact]{titlesec}         
\titlespacing{\section}{8pt}{8pt}{8pt} 
\AtBeginDocument{
  \setlength\abovedisplayskip{2pt}
  \setlength\belowdisplayskip{2pt}}

\newtheorem{remark}{Remark}

\newcommand{\iu}{{j\mkern1mu}}
\usepackage{soul}
\allowdisplaybreaks
\def\BibTeX{{\rm B\kern-.05em{\sc i\kern-.025em b}\kern-.08em
    T\kern-.1667em\lower.7ex\hbox{E}\kern-.125emX}}
\begin{document}

\title{
Unified Control Scheme for Optimal Allocation of GFM and GFL Inverters in Power Networks
}
\author{Sushobhan Chatterjee\orcidA{} and Sijia Geng\orcidB{}
\thanks{The authors are with the Department of Electrical and Computer Engineering, Johns Hopkins University. \texttt{(email: schatt21@jhu.edu; sgeng@jhu.edu)}.}%
}

\maketitle

\begin{abstract}
With the rapid adoption of emerging inverter-based resources, it is crucial to understand their dynamic interactions across the network and ensure stability.
This paper proposes a systematic and efficient method to determine the optimal allocation of grid-forming and grid-following inverters in power networks. 
The approach leverages a novel unified grid-forming/following inverter control and formulates an optimization problem to ensure stability and maximal energy dissipation during transient periods. An iterative algorithm is developed to solve the optimization problem. The resulting optimal droop gains for the unified inverters provide insights into the needs for grid-forming and grid-following resources in the network. A three-bus system is used to demonstrate the optimality and dynamic performance of the proposed method.
\end{abstract}

\begin{IEEEkeywords}
Inverter-based resources, unified control, optimization, stability, Lyapunov equation, iterative algorithm
\end{IEEEkeywords}

\section{Introduction} \label{sec:intro}
Inverter-based resources (IBRs) of heterogeneous natures will be prevalent in future power systems, which demands proper coordination of their control schemes to ensure system-level stability. One major challenge lies in the lack of clarity of inverter models and their unknown interactions with and through the network. 
Due to the high complexity, current practices perform case-by-case analysis in determining the appropriate integration level and mix of IBRs in the network and fail to provide general statements. 
Moving forward, systematic and computationally efficient tools need to be developed.
 
The dynamics of IBRs as seen from the grid are dominated by the enforced control law \cite{milano2018foundations}. Control strategies are typically classified as grid-following (GFL) or grid-forming (GFM) \cite{li2022revisiting}. There are numerous ways of implementing GFM \cite{zhong2010synchronverters,chandorkar1993control,torres2015synchronization,sinha2015uncovering} and GFL \cite{yang2022bifurcations,pal2023large,prakash2023modeling} inverters and the classification is not exhaustive \cite{geng2022unified}. The majority of IBRs currently in use are of the GFL type. These inverters normally rely on phase-locked loops (PLLs) to synchronize with the grid and deliver specified amounts of active and reactive power. 
In contrast, GFM inverters normally act as controllable voltage sources behind a coupling reactance, resembling the behavior of synchronous generators. 
This capability allows them to perform black-start operations (establishing voltage on a de-energized network) and support the voltage in weak networks. Despite these advantages, the dynamic interactions between GFL and GFM control strategies under various network conditions remain an area of ongoing research \cite{matevosyan2022grid, geng2022unified}. 

This paper aims to develop a method to provide insights on the needs for ``forming'' and ``following'' resources across the network in order to guide the placement of various IBRs. 
The method uses an optimization problem to formulate the stability characterization and dynamic performance of the networked dynamical system.
We avoid the complexity of accounting for all possible combinatorial allocations of various types of IBRs within the network, instead adopting a systematic approach to decision-making by leveraging a novel unified grid-forming/following inverter \cite{geng2022unified, geng2022dynamic}. This unified control scheme seamlessly integrates elements of grid-forming (i.e., P-$\omega$ and V-Q droop characteristics) and grid-following (i.e., PLL) controls. The magnitude of the droop gains represents the strength of the ``forming'' capability versus the ``following'' feature. 
We optimize a single control parameter for each unified inverter, that is, the P-$\omega$ droop gain, to ensure stability and optimize system performance by extracting the largest amount of energy (in the sense of dissipativity theory \cite{veselic2011damped}) during transients. {\color{black} An iterative algorithm is proposed to solve the non-convex optimization problem.} 
The optimal droop gains provide an understanding of the network needs for GFM and GFL capability. Motivated by these considerations, heuristics can be designed to place various GFM and GFL resources in the network accordingly. 

The remainder of the paper is organized as follows: Section~\ref{sec:prelims} presents the preliminaries and Section~\ref{sec:prob} gives the problem formulation. An iterative algorithm is proposed in Section~\ref{sec:algo} to solve the optimization problem. Numerical results are discussed in Section~\ref{sec:res} and conclusions are given in Section~\ref{sec:conclusion}.

\section{Preliminaries} \label{sec:prelims}
In this section, we present the preliminaries on notations, power system modeling, and inverter models.
\subsection{Notations}
Let $\mathbb{R}^n$ and $\mathbb{C}^n$ represent the spaces of $n$-dimensional real and complex vectors, respectively, and let $\mathbb{R}^{n \times n}$ denote the space of $n \times n$ real matrices. Define $I_n$ as the identity matrix of size $n$. For a scalar $z \in \mathbb{C}$, $|z|$ represents its absolute value, and $\Re(z)$ denotes the real part of a complex number $z$, while $\Im(z)$ represents its imaginary part. For a vector $x\in \mathbb{C}^n$, $x_i$ indicates its $i$th component and $x^T$ stands for its transpose. Unless stated otherwise, the notation $\|x\|$ denotes the 2-norm of the vector $x$. $x \succeq y$ is defined as component-wise inequality between $x$ and $y$, that is, $x_i \geq y_i$ for every index $i$. 
Lastly, for a matrix $P \in \mathbb{R}^{n \times n}$, $\mu_i(P)$ refers to its $i$th eigenvalue, $\Lambda(P)$ represents the set of its eigenvalues, and $\rho(P)$ denotes its spectral radius, i.e., $\max_{i}(|\mu_i(P)|)$. $P \succ 0$ and $P \succcurlyeq 0$ denote positive-definite and positive-semidefinite matrices, respectively. The $(i, j)$th element of $P$ is written as $[P]_{ij}$, and $\lVert P \rVert$ is a matrix norm on $(n \times n)$ dimensional matrix space.

\subsection{Network Modeling}\label{sec:network}
\begin{figure}[ht!]
\centerline{\includegraphics[scale=1]{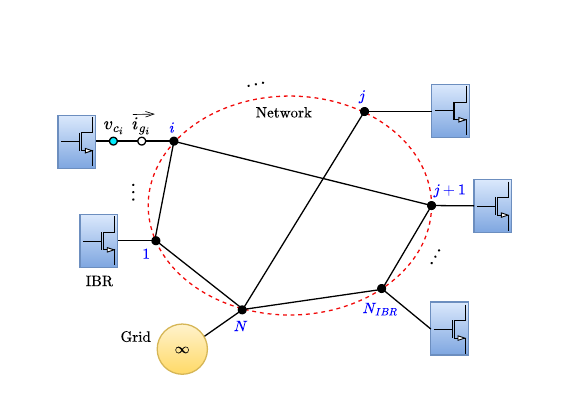}}
    \vspace{-3ex}
    \caption{Structure of a network-reduced inverter-based power network.}
    \label{fig:nbus} 
\end{figure}

This paper considers network-reduced inverter-based power systems. Without loss of generality, assume there are in total $N$ buses. The first $N_\text{IBR}$ buses are connected with inverters and the last bus is the slack bus which provides angle reference. Fig. \ref{fig:nbus} illustrates the network structure. \vspace{-9ex}

A global rotating $DQ$-frame transforms sinusoidal signals into nearly constant values at the steady state. The angular velocity of this frame is defined as $\omega_{DQ} \omega_b$ rad/s, where $\omega_b$ is the base frequency (e.g., $2\pi \times 50$ rad/s in Europe and most of Asia, or $2\pi \times 60$ rad/s in North America) and $\omega_{DQ}$ is the per-unit frequency of the $DQ$-frame (typically taken as $1$). The voltages and currents of the network are expressed in the $DQ$-frame as $v_t = v_{t,D} + jv_{t,Q}$ and $i_t = i_{t,D} + ji_{t,Q}$.
For each inverter, defining their local $dq$-frame is common, which rotates at an angular frequency of $(\Delta \omega + \omega_0)\omega_b$ rad/s. Here, $\Delta \omega$ is the per-unit frequency deviation, determined by the inverter's control mechanism (e.g., PLL or droop control). Terminal voltages and currents are represented in the local $dq$-frame as $v_t = v_{t,d} + jv_{t,q}$ and $i_t = i_{t,d} + ji_{t,q}$.
To unify power system representation for analysis purposes, local $dq$-frame variables are transformed into the global $DQ$-frame through, 
\begin{equation}
\begin{aligned}
\begin{bmatrix} v_{t,D} \\ v_{t,Q} \end{bmatrix} &= \mathcal{R}(\theta) \begin{bmatrix} v_{t,d} \\ v_{t,q} \end{bmatrix}, \quad 
\begin{bmatrix} i_{t,D} \\ i_{t,Q} \end{bmatrix} = \mathcal{R}(\theta) \begin{bmatrix} i_{t,d} \\ i_{t,q} \end{bmatrix},
\end{aligned}    
\end{equation}
where $\theta$ is the angle between the local and global frames and $\mathcal{R}(\theta)$ is the rotation matrix defined as, 
\begin{equation}
    \mathcal{R}(\theta) = \begin{bmatrix} \cos(\theta) & -\sin(\theta) \\ \sin(\theta) & \cos(\theta) \end{bmatrix}.
\end{equation}

Transmission lines connect the buses. We use the commonly adopted static line model to describe power flow equations.
The current flowing from an inverter located at the $k$th bus relates to the neighboring buses' voltages and line impedances as, 
\begin{equation} \label{sys:line}
    \begin{aligned}
        i_{g_k,D} & = \overset{n}{\underset{l=1}{\mathlarger{\Sigma}}} \Big[\Re\{y_{kl}\} \big(v_{c_k,D} \!-\! v_{c_l,D}\big) \!-\! \Im\{y_{kl}\} \big(v_{c_k,Q} \!-\! v_{c_l,Q}
        \big) \Big], 
        \\
        i_{g_k,Q} & = \overset{n}{\underset{l=1}{\mathlarger{\Sigma}}} \Big[\Re\{y_{kl}\} \big(v_{c_k,Q} \!-\! v_{c_l,Q}\big) \!+\! \Im\{y_{kl}\} \big(v_{c_k,D}\!-\! v_{c_l,D}
        \big) \Big]. 
    \end{aligned}
\end{equation}

An object-oriented method for assembling the components of a power system is introduced in \cite{hiskens2001systematic}. Each component is individually modeled using its differential-algebraic equations (DAEs), enabling symbolic computation of all partial derivatives. These equations and their derivatives are then combined to construct the DAE model to describe the entire power system in the form of,
\begin{equation} \label{eqn:app1}
    \begin{aligned}
        \dot{x} &\ = f(x,y,p)
        \\
        0 &\ = g(x,y,p),
    \end{aligned}
\end{equation}
where $x\in\mathbb{R}^n$ are the dynamic states, $y\in\mathbb{R}^m$ are the algebraic states, and $p\in\mathbb{R}^l$ are the parameters. At an equilibrium, $f(x^{*},y^{*},p)=0,\ g(x^{*},y^{*},p)=0$.
The calculation also constructs the Jacobian matrices symbolically and the linearized DAE model,
\begin{equation}
    \begin{bmatrix}
        \dot{\Delta x} \\
        0
    \end{bmatrix} = \begin{bmatrix}
        \frac{\partial f}{\partial x} & \frac{\partial f}{\partial y} \\
        \frac{\partial g}{\partial x} & \frac{\partial g}{\partial y}
    \end{bmatrix} \begin{bmatrix}
        \Delta x \\
        \Delta y
    \end{bmatrix}.
\end{equation}
To evaluate the small-disturbance stability of the system around an equilibrium ($x^{*},y^{*},p$), the algebraic states $y$ are eliminated, resulting in an effective form,
\begin{equation} \label{eqn:88}
    \dot{\Delta x} = \Bigg(\frac{\partial f}{\partial x} - \frac{\partial f}{\partial y} \bigg(\frac{\partial g}{\partial y}\bigg)^{-1} \frac{\partial g}{\partial x} \Bigg)\Bigg|_{*} \Delta x \triangleq A_\text{eff}(x^{*},y^{*},p) \Delta x. 
\end{equation}
The eigenspectrum $\Lambda (A_\text{eff})$ determines the small-disturbance stability of the system. 

\subsection{Model of the Unified Grid-Forming/Following Inverter}\label{sys:uni}
\begin{figure}[ht!]
    \hspace{-3ex} \centerline{\includegraphics[scale=0.67]{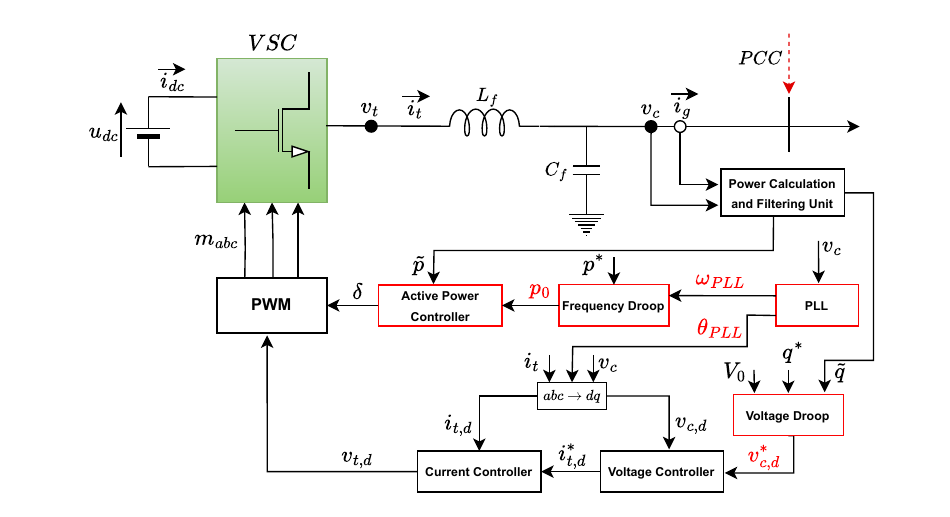}}
    \caption{Control blocks of the unified grid-forming/following inverter.}
    \label{fig:setup_uni} 
\end{figure}

\begin{figure}[ht!]
    \hspace{0ex} \centerline{\includegraphics[scale=0.8]{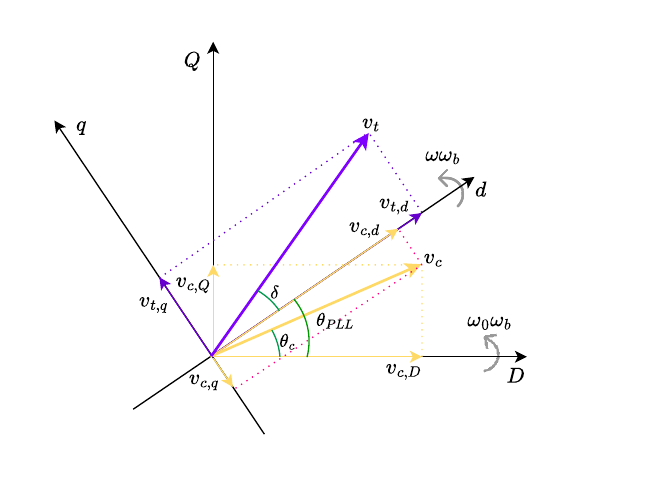}}
    \caption{Phasor diagram and reference frames of the unified inverter.}
    \label{fig:uni_ph} 
\end{figure}

This section provides an overview of the unified inverter control scheme \cite{geng2022unified} that serves as the basis for the proposed method. 
Fig. \ref{fig:setup_uni} illustrates the control blocks which integrates characteristics from both grid-forming and grid-following inverters. For the detailed implementation of the control blocks, refer to Fig. \ref{fig:control_uni} and the equations in the Appendix. 
The controller regulates the voltage magnitude $v_c$ at the point of common coupling (PCC) and the active power $p$ supplied to the grid. 
This setup includes a voltage source converter (VSC) with an output {LC} filter, described by equations \eqref{eqn:su9}-\eqref{eqn:su12}.
A phase-locked loop ({PLL}) \eqref{eqn:su6} generates the angle \eqref{eqn:su7} and frequency \eqref{eqn:au1}, \eqref{eqn:au5} that are required to transform three-phase voltage and current signals into {$dq$}-frame quantities. Each control loop operates in a local {$dq$} reference frame, utilizing Park’s transformation, as described in \eqref{eqn:au8}-\eqref{eqn:au12} and \eqref{eqn:au15}-\eqref{eqn:au16}, and illustrated in Fig. \ref{fig:uni_ph}. 
The outer loop includes a $d$-axis voltage controller ({VC}) governed by \eqref{eqn:su3}, and an active power controller ({APC}) governed by \eqref{eqn:su4}-\eqref{eqn:su5}. A notable feature of the controller is its ability to regulate active power by modulating the phase angle across the filter inductance. Reference signal generated by the voltage-based outer loop, as defined in \eqref{eqn:au13}, is passed onto the inner $d$-axis current control ({CC}) loop, which is described by \eqref{eqn:su8} and \eqref{eqn:au14}. On the other hand, the output of the power-based outer loop is directly used to control the load angle $\delta$ given by \eqref{eqn:au4}. The DC power balance is captured in \eqref{eqn:au17}.  

Furthermore, unified inverter controls equipped with droop characteristics enable frequency and voltage regulation. The active power-frequency (P-$\omega$) droop mechanism given by \eqref{eqn:au2} provides the load angle setpoint, while the voltage-reactive power (V-Q) droop mechanism given by \eqref{eqn:au3} determines the voltage setpoint. These droop mechanisms use filtered power measurements \eqref{eqn:su1}-\eqref{eqn:su2} as inputs. 

Robustness of the unified inverter control has been validated through experiments on a hardware testbed \cite{surprenant2011phase} and simulations \cite{geng2022unified, geng2022dynamic}. 
It can achieve autonomous islanding and reconnection in microgrid settings without requiring control switching. 

\section{Problem Formulation} \label{sec:prob}
    Considering the network-reduced inverter-based power systems (Section~\ref{sec:network}). Assume that each node in the network is open to connection to an IBR with a unified controller (Section~\ref{sys:uni}). We aim to determine the optimal profile of the P-$\omega$ droop gains for the unified controllers across the network, such that, 
\begin{enumerate}
    \item The system is small-disturbance stable, i.e., $\forall\ i$, $\Re \big\{\mu_i(A_{\text{eff}})\big\}<0$, and,
   \item Dynamic performance is optimized by minimizing $\int^{\infty}_0 x(t)^TQx(t) dt \geq 0$.
    \end{enumerate}

To achieve the goals, we consider a formulation by introducing the "Lyapunov trace", which is based on the principles of asymptotic stability and system dissipativity \cite[Chapter 21]{veselic2011damped}. This formulation ensures optimal damping to maximize energy extraction from the system, specifically aiming to minimize the stored energy represented by $\int^{\infty}_0 x(t)^TQx(t) dt \geq 0$.
The formulation is as follows. 
\begin{mini*}|1|
 {\substack{\mathbf{K_P}, P,x^{*},y^{*}}}{\text{Trace}\big(PS\big)}{}{}
 \label{eqn:P2}
 \addConstraint{\tilde{A}_{\text{eff}}^{T}P + P\tilde{A}_{\text{eff}} = -Q}
 \addConstraint{P \succ 0} \tag{$A$}
 \addConstraint{\mathbf{\underline{K}_P} \preceq \mathbf{K_P} \preceq \mathbf{\overline{K}_P}} 
 \addConstraint{\tilde{A}_{\text{eff}} = A_{\text{eff}}(x^{*},y^{*},p_0,\mathbf{K_P})}
 \addConstraint{f(x^{*},y^{*},p_0,\mathbf{K_P})=0}
 \addConstraint{g(x^{*},y^{*},p_0,\mathbf{K_P})=0}.
\end{mini*}
In Problem \ref{eqn:P2}, decision variables $\mathbf{K_P} \in \mathbb{R}_{\geq 0}^{N_\text{IBR}}$ denote the vector of P-$\omega$ droop gains, which are box-constrained by $\mathbf{\underline{K}_P}$ and $\mathbf{\overline{K}_P}$, following engineering considerations. The rest of the parameters, $p_0:= p \setminus \{K_{Pi}\}$, remain constant. 
The parameters $Q$ and $S$ are predefined symmetric positive definite matrices. Typically, we set $Q = I_n$, where $n$ is the order of $A_{\text{eff}}$, to ensure equal weighting for all states in terms of energy extraction. However, a more informed choice for $Q$ can be made, for example, to target and penalize specific states that contribute to the slowest modes as determined by the \textit{modal analysis}. A notable selection for the weight matrix $S$ is given by $S = I_n / (2n)$. This choice aligns with the Lebesgue measure which equally weights all states when minimizing the system's energy. 

The first two constraints in Problem \ref{eqn:P2} impose a stability guarantee through the Lyapunov equation. Specifically, these constraints guarantee that the linearized system $A_\text{eff}$ is \textit{globally} uniformly asymptotically stable, provided there exists a $P \succ 0$ for a given $Q \succ 0$ \cite{boyd2008notes}. This requirement also ensures that $P$ is uniquely determined and that the original nonlinear system achieves \textit{local} uniform asymptotic stability \cite[Theorem 4.6]{murray2017mathematical}. The subsequent two constraints enforce parameter bounds and maintain the structural integrity of the effective system matrix. The last two constraints establish the relationship between $x^{*}$ and $y^{*}$ with $\mathbf{K_P}$ through the nonlinear equilibrium equations.

Directly solving Problem \ref{eqn:P2} presents several challenges. First, the nonlinear constraints defining the equilibrium condition add complexity. Second, the Lyapunov equation becomes nonlinear because both $\tilde{A}_{\text{eff}}$ and \( P \) include decision variables. To address these difficulties, the next section introduces an iterative algorithm that exploits the structural properties of the \(\tilde{A}_{\text{eff}}\) matrix.

\section{Algorithm} \label{sec:algo} 
 In this section, we propose an iterative algorithm to solve the Problem \ref{eqn:P2}. We decouple the updates of the equilibrium point $x^{*},\ y^{*}$ and the updates of droop parameters $\mathbf{K_P}$.
 At each iteration $k$, Algorithm \eqref{algo:nonlinopt} first solve the Problem \ref{eqn:P3} using the previous equilibrium values $x^{*}[k-1]$ and $y^{*}[k-1]$.
 \begin{mini*}|1|
 {\substack{\mathbf{K_P},\ P}}
 {\text{Trace}\big(PS\big)}{}{}
 \label{eqn:P3}
 \addConstraint{\tilde{A}_{\text{eff}}^{T}P + P\tilde{A}_{\text{eff}} = -Q}
 \addConstraint{P \succ 0} \tag{$B$}
 \addConstraint{\mathbf{\underline{K}_P} \preceq \mathbf{K_P} \preceq \mathbf{\overline{K}_P}} 
 \addConstraint{\tilde{A}_{\text{eff}} = A_{\text{eff}}(x^{*},y^{*},p_0,\mathbf{K_P})}.
\end{mini*}

 \begin{remark}
    Note that the entries of system matrix $A_\text{eff}$ have a special structure that is affine in the non-negative decision variables $\mathbf{K_P}$. Furthermore, the Lyapunov equation is a bilinear matrix inequality (BMI) in $\tilde{A}_\text{eff}$ and $P$.
\end{remark}

 After calculating the optimal gains $\mathbf{K_P^{\text{O}}}$, these values are passed onto the next stage, where it is verified whether the P-$\omega$ droop gains $\mathbf{K_P}[k]$, together with the previous equilibrium point $x^{*}[k-1]$ and $y^{*}[k-1]$, satisfy the equilibrium condition,
 \begin{equation} \label{eqn:sc}
    R[k] = \lVert r_1[k] \rVert_1 + \lVert r_2[k] \rVert_1 \leq \epsilon,
\end{equation}
where,
 \begin{equation} \label{eqn:eq}
    \begin{aligned}
    r_1[k] &\ = f\big(x^{*}[k-1],y^{*}[k-1],p_0,\mathbf{K_P}[k]\big),
    \\
    r_2[k] &\ = g\big(x^{*}[k-1],y^{*}[k-1],p_0,\mathbf{K_P}[k]\big).
    \end{aligned}
\end{equation}
$R[k]$ is the {residual error} calculated at each step $k$ and $\epsilon > 0$ is a chosen small tolerance value.

Algorithm \ref{algo:nonlinopt} is terminated when the condition is satisfied.
 If any violations are detected by comparing the residual error $R[k]$ to the tolerance, the algorithm moves on to calculate a new equilibrium point $x^{*}$, $y^{*}$ at the $k$th step by solving the nonlinear DAE equations with the obtained $\mathbf{K_P}[k]$, 
\begin{equation} \label{eqn:eqm}
    \begin{aligned}
        f(x^{*},y^{*},p_0,\mathbf{K_P}[k]) &\ = 0,
        \\
        g(x^{*},y^{*},p_0,\mathbf{K_P}[k]) &\ = 0.
    \end{aligned}
\end{equation}
This process repeats until the stopping criteria are met, at which point the algorithm terminates. 
 
\begin{algorithm}[H]
  \caption{Iterative algorithm to compute optimal P-$\omega$ droop gains for unified IBRs in an N-bus network.}
  \label{algo:nonlinopt}
  \begin{algorithmic}
    \State \textbf{Given:} Bounds $\rightarrow$ $\big\{\mathbf{\underline{K}_{P}}, \mathbf{\overline{K}_{P}}\big\}$, Parameters $\rightarrow$ $p_0$, $\epsilon > 0$, $S$, $Q$, $x^{*}[0]$, $y^{*}[0]$ 
    \State \textbf{Initialization:} $converged = 0$, $k = 1$ 
    \State \textbf{Iterate:} 
    \While{$converged==0$}
    \State \textbf{Optimization:}
    \State Solve Problem \ref{eqn:P3} for optimal droop gains $\mathbf{K_P^{\text{O}}}$, using $x^{*}[k-1]$ and $y^{*}[k-1]$
    \State Fix $\mathbf{K_P}[k] = \mathbf{K_P^{\text{O}}}$
    \State \textbf{Equilibrium Check:}
    \State Compute $R[k]$ as per \eqref{eqn:sc}, using $x^{*}[k-1]$, $y^{*}[k-1]$ and $\mathbf{K_P}[k]$ 
    \If{$R[k] \leq \epsilon$}
        \State $converged = 1$
    \Else
        \State $converged = 0$
    \EndIf
    \State \textbf{Equilibrium Point Computation:}
    \State Evaluate $x^{*}[k]$ and $y^{*}[k]$ as per \eqref{eqn:eqm}, using $\mathbf{K_P}[k]$ 
    \State Fix $k \rightarrow k+1$
    \EndWhile  
  \end{algorithmic}
 \end{algorithm}

\begin{remark}
Algorithm \eqref{algo:nonlinopt} is started at $k=0$ with the nominal stable equilibrium point ($x^{*}$, $y^{*}$) corresponding to some nominal $\mathbf{K_P}$.
\end{remark}

\section{Numerical Results} \label{sec:res}
\begin{figure}[ht!]
\centerline{\includegraphics[scale=0.75]{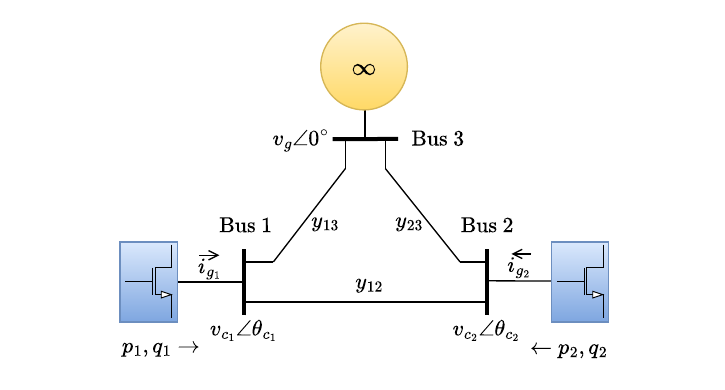}}
    \caption{Schematic of a three-bus power system.}
    \label{fig:3bus} 
\end{figure}

This section presents numerical results based on a three-bus system as depicted in Fig. \ref{fig:3bus}. Inverters are connected to buses 1 and 2, with bus 3 serving as the slack bus, providing the angle reference. The buses are interconnected through static transmission lines. 
We investigate the performance of the iterative algorithm, focusing on its convergence and validity. 
{\color{black} The optimization problems were solved on an Apple MacBook Pro (M3 chip with 11 cores) using MATLAB R2024a. The YALMIP global BMIBNB method \cite{Lofberg2004} was employed, utilizing \textit{fmincon} as the upper-level solver and MOSEK as the lower-level solver.}
Finally, GFM and GFL inverters are connected to the network to replace the unified inverters using heuristics guided by the optimal droop gains. System stability and dynamic performance are evaluated. Nominal parameters from the literature are used without additional tuning to demonstrate the efficiency of the approach.

\subsection{Algorithm Performance}
Assume two unified inverters are connected at buses 1 and 2, whose parameters are provided in Tables \eqref{tab:t1} and \eqref{tab:t2}. Unless specified otherwise, all parameters are expressed in per unit. The iterative algorithm is executed to optimize the P-$\omega$ droop gains of the unified inverters. The nominal equilibrium points $x^{*}[0]$, $y^{*}[0]$ used for initializing the algorithm were generated with nominal droop gains of $K_{P1} = K_{P2} = 10$.
\begin{table}[ht!]
    \centering
  \setlength\tabcolsep{0.8pt}
  \setlength\extrarowheight{2pt}
  \caption{Parameters of the unified inverters at Bus 1 and 2 (in p.u.)}
    \begin{tabular}{cccccccccc}
      \Xhline{2\arrayrulewidth} 
      $V_0$ & $\omega_0$ & $L_f$ & $C_f$ & $\omega_b$ & $K_Q$ & $K^P_{PC}$ &  $K^I_{PC}$ & $K^P_{VC}$ & $K^I_{VC}$ \vspace{0.8ex}\\
      \hline
      1 & 1 & 0.1 & 0.3 & 120$\pi$ & 0.05 & 0.23 & 0.6 & 1 & 2 \\
      \hhline{|==========|}
       $\omega_{pc}$ & $K^F_{VC}$ & $K^P_{CC}$ & $K^I_{CC}$ & $K^F_{CC}$ & $K^P_{PLL}$ & $K^I_{PLL}$ & $q^{*}$ & $\omega_{qc}$ & \vspace{0.8ex}\\ 
       \hline
       332.8 rad/s & 1 & 1 & 2 & 0 & 0.2 & 5 & 0.25 & 732.8 rad/s & \\ 
      \Xhline{2\arrayrulewidth} 
    \end{tabular}
    \label{tab:t1}
\end{table}

\begin{table}[ht!]
  \setlength\tabcolsep{5pt}
  \setlength\extrarowheight{2pt}
  \centering
  \caption{Power set-points and bounds of the unified inverters (in p.u.)}
    \begin{tabular}{cccc}
      \Xhline{2\arrayrulewidth} 
      Location of IBR & $p^{*}$ & $\mathbf{\underline{K}_P}$ & $\mathbf{\overline{K}_P}$ \\
      \hline
      Bus 1 & 0.8 & 0 & 1200 \\
      Bus 2 & 0.2 & 0 & 1200 \\
      \Xhline{2\arrayrulewidth} 
    \end{tabular}
    \label{tab:t2}
\end{table}

The parameters of the transmission lines are given in Table~\eqref{tab:lin1}.
\begin{table}[ht!]
  \setlength\tabcolsep{5pt}
  \setlength\extrarowheight{2pt}
  \centering
  \caption{Transmission line data for the three-bus network (in p.u.)}
    \begin{tabular}{ccc}
      \Xhline{2\arrayrulewidth} 
      $y_{12}$ & $y_{13}$ & $y_{23}$ \\
      \hline
      0.0917 - $\iu$3.0275 & 3.4910 - $\iu$12.7422 & 3.4910 - $\iu$12.7422 \\
      \Xhline{2\arrayrulewidth} 
    \end{tabular}
    \label{tab:lin1}
\end{table}

\begin{figure}[ht!]
    \hspace{-1ex} \centerline{\includegraphics[scale=0.37]{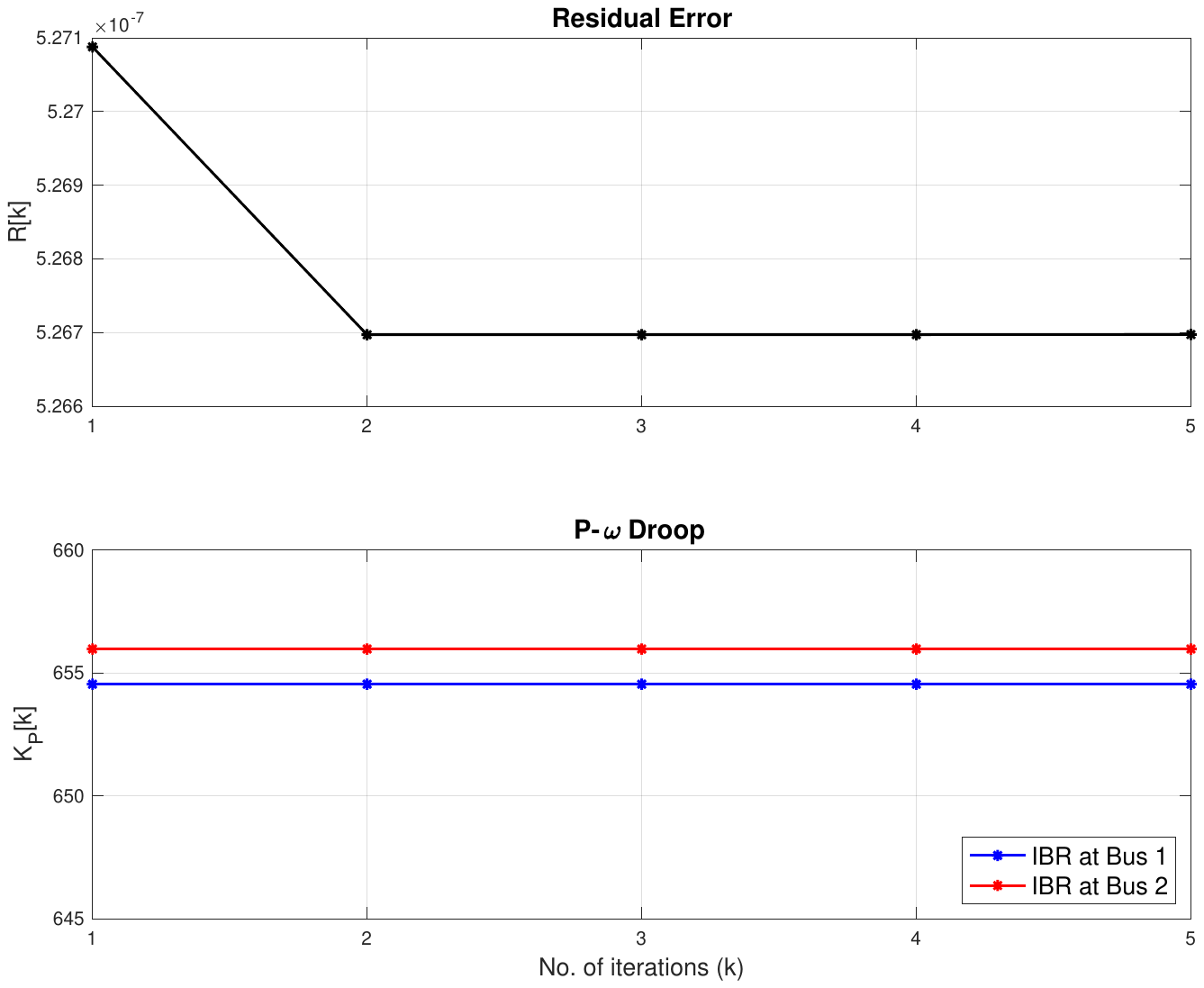}}
    \caption{Performance of the iterative algorithm (Base impedance).}
    \label{fig:alg1} 
\end{figure}

Fig. \ref{fig:alg1} shows the algorithm's performance. It requires only two iterations for the algorithm to converge to the final values where the residual error drops to $5.267 \times 10^{-7}$. The P-$\omega$ droop gains achieve optimality at $K^O_{P1} = 654.546$ and $K^O_{P2} = 655.978$, respectively. Note that the obtained solution might not necessarily be globally optimal for Problem~\ref{eqn:P2}. However, numerical scanning of the feasible region verifies the global optimality of the solution in this example.

It can be observed that the optimal P-$\omega$ droop gains of the two inverters are close to each other due to the symmetry of the three-bus network. The droop gain of the inverter at bus 2 is slightly higher than that at bus 1. This is because the bus 2 inverter operates at a lower active power setpoint, thereby requiring less stringent power output regulation.
In contrast, bus 1 offers greater flexibility for control adjustments with its smaller droop.

\subsection{Standard GFM and GFL Models}
We use the optimized P-$\omega$ droop gains of the unified inverters as guidance to support the allocation of GFL IBRs (which exhibit zero droop behavior and use a PLL for synchronization) and GFM IBRs (which display positive $\omega$-P droop behavior). Two standard GFL and GFM inverter models as described below are used as an example to demonstrate the idea.

\subsubsection{Grid-Following (GFL) Inverter} \label{sys:gfl}
\begin{figure}[ht!]
    \hspace{4.5ex}\rightline{\includegraphics[scale=0.7]{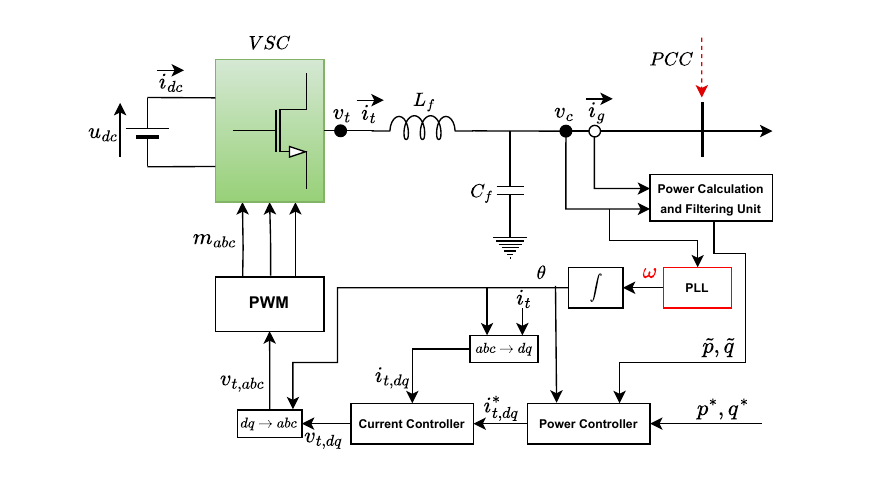}}
    \caption{Control blocks of the GFL inverter.}
    \label{fig:setup_gfl} 
\end{figure}

The control blocks of the GFL inverter are illustrated in Fig.~\ref{fig:setup_gfl}. The detailed DAE model, control implementation (Fig. \ref{fig:control_gfl}), along with the nomenclature and reference frames (Fig. \ref{fig:gfl_ph}) are provided in the Appendix. Its control scheme employs a hierarchical structure with an outer power control loop and an inner current control loop. Synchronization is achieved by the PLL.

\subsubsection{Grid-Forming (GFM) Inverter} \label{sys:gfm}
\begin{figure}[ht!]
    \hspace{-3ex} \centerline{\includegraphics[scale=0.7]{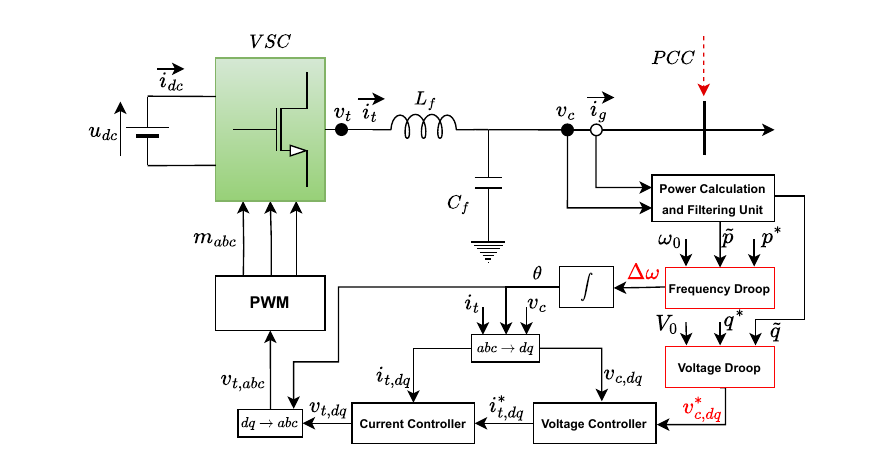}}
    \caption{Control blocks of the GFM inverter.}
    \label{fig:setup_gfm} 
\end{figure}

The control architecture of the GFM inverter is depicted in Fig.~\ref{fig:setup_gfm}. Detailed information on the DAE model, control implementation (Fig. \ref{fig:control_gfm}), and the associated nomenclature and reference frames (Fig. \ref{fig:gfm_ph}) can be found in the Appendix. The control scheme adopts a nested structure, consisting of an outer voltage control loop and an inner current control loop, with synchronization achieved via droop characteristics.

\begin{remark} \label{rem:1}
The unified inverter employs P-$\omega$ droop control for active power regulation, whereas the GFM inverter adopts $\omega$-P droop for the same purpose. An inverse relationship exists between these two droop mechanisms, as evidenced by \eqref{eqn:am1.5} and \eqref{eqn:au2}. Additionally, the droop gain $K_P$ for GFM inverters is represented as a percentage, while for the unified inverter, it is given in per unit; these two formats of droop parameters can be readily converted between each other.
\end{remark}

\subsection{Evaluation of Dynamic Performance}
We evaluate the dynamic performance of the three-bus power system featuring various mixes of GFM, GFL, and unified inverters at its buses. The configurations consist: Unified-only, GFL-only, GFM-only, and mixed GFM-GFL setups. To assess the robustness of our approach under varying line impedances, we analyze three scenarios with different impedance values. The values given in Table \ref{tab:lin1} serve as the base case, with low and high impedances subsequently evaluated and compared. The power set-points and bounds remain the same as in Table \ref{tab:t2}.

\subsubsection{Case 1: Base impedance network}
\begin{figure}[ht!]
    \hspace{0ex} \centerline{\includegraphics[scale=0.37]{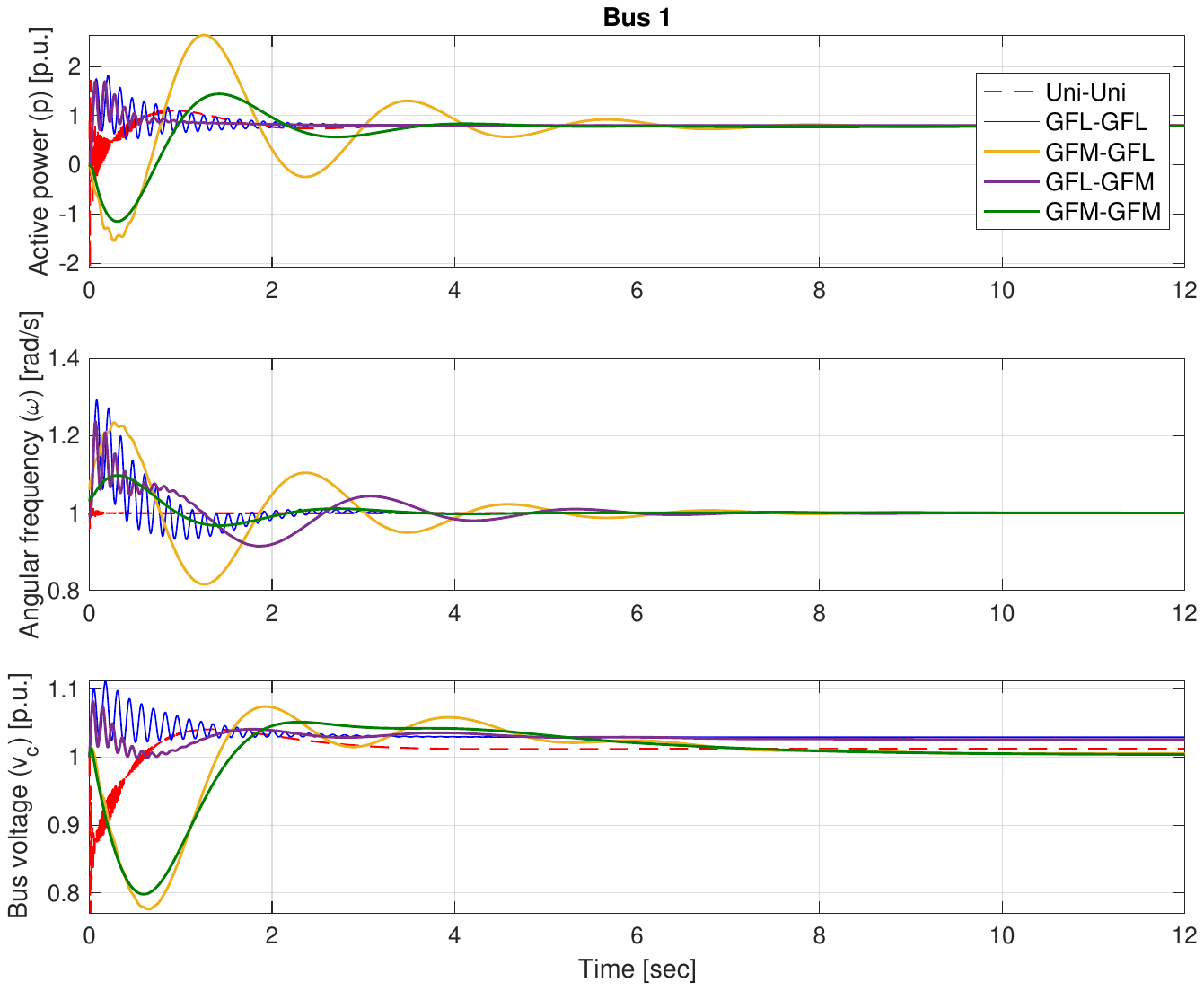}}
    \caption{Trajectories of active power, angular frequency, and voltage at Bus 1 (Base impedance).}
    \label{fig:ca1} 
\end{figure}

\begin{figure}[ht!]
    \hspace{0ex} \centerline{\includegraphics[scale=0.37]{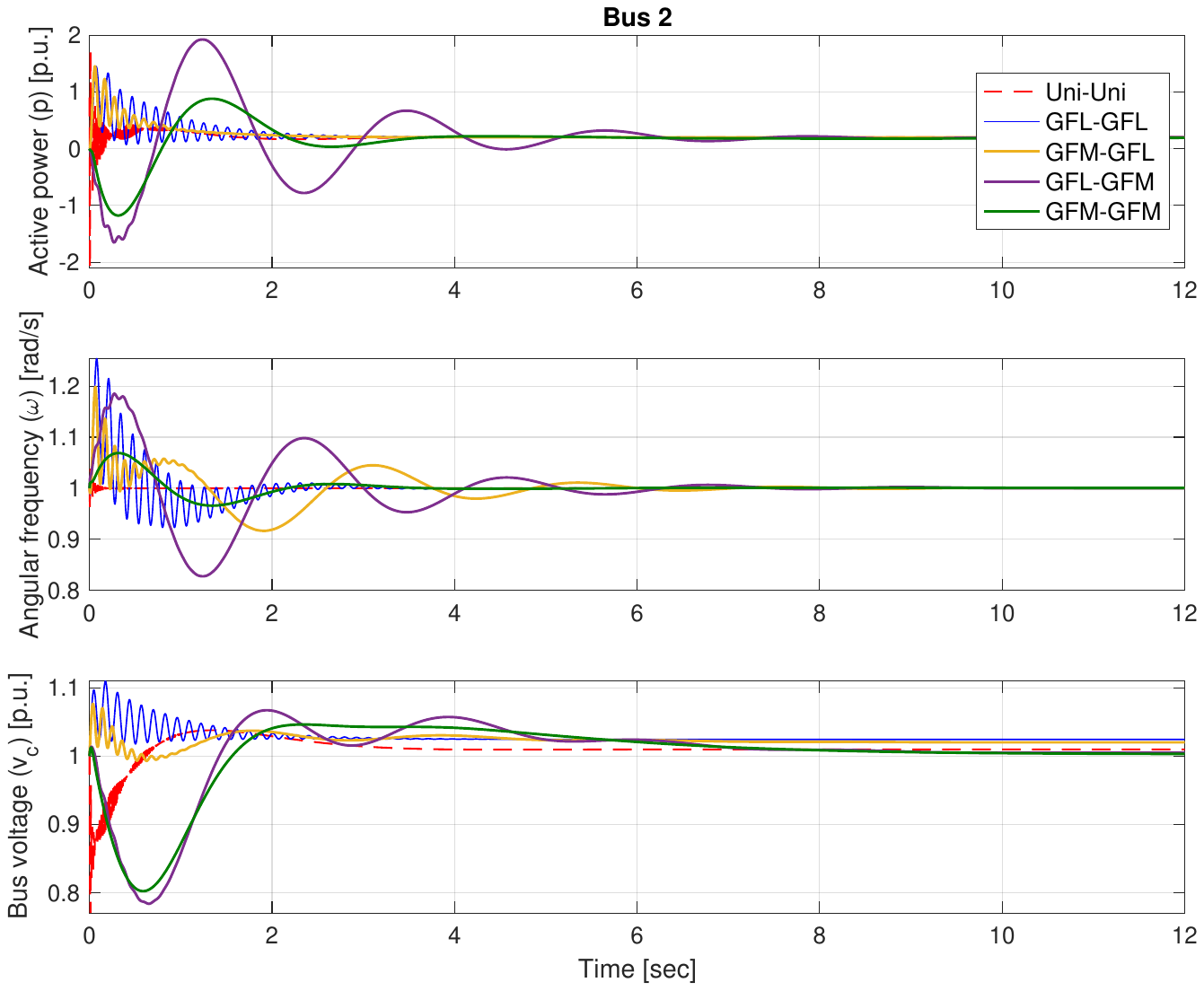}}
    \caption{Trajectories of active power, angular frequency, and voltage at Bus 2 (Base impedance).}
    \label{fig:ca2} 
\end{figure}

Impedance values given in Table \ref{tab:lin1} produce the benchmark results as shown in Figs. \ref{fig:ca1} and \ref{fig:ca2}, corresponding to buses 1 and 2, respectively. Recall that in this case, the iterative algorithm gives optimal droop gains as $\mathbf{K_P} = [654.546, 655.979]$, which are { close to each other. Correspondingly, the mixed GFM/GFL cases perform the worst in this example. Other than the trajectory of the optimal unified case, the GFM-GFM and GFL-GFL systems give similar performances in terms of settling time. However, they perform poorly with respect to undershoots and oscillations, respectively. } 

\begin{remark}
   Note that the dynamic performance metrics, including over/undershoots, oscillations, and decay rates, have not been tuned. While performance can be improved with proper tuning, it's not the focus of demonstrating the idea. 
\end{remark}

\subsubsection{Case 2: Low-impedance network}
To consider a stronger grid case, the line impedances are reduced {\color{black}by $27\%$, $23\%$ and $23\%$ (relative to the base values) for lines 1-2, 1-3 and 2-3, respectively}, as given in Table~\ref{tab:lin2}. 
\begin{table}[ht!]
  \setlength\tabcolsep{5pt}
  \setlength\extrarowheight{2pt}
  \centering
  \caption{Transmission line data for 3-bus network (in per units)}
    \begin{tabular}{ccc}
      \Xhline{2\arrayrulewidth} 
      $y_{12}$ & $y_{13}$ & $y_{23}$ \\
      \hline
      0.1387 - $\iu$4.1620 & 4.717 - $\iu$16.5093 & 4.717 - $\iu$16.5093 \\
      \Xhline{2\arrayrulewidth} 
    \end{tabular}
    \label{tab:lin2}
\end{table}

\begin{figure}[ht!]
    \hspace{0ex} \centerline{\includegraphics[scale=0.37]{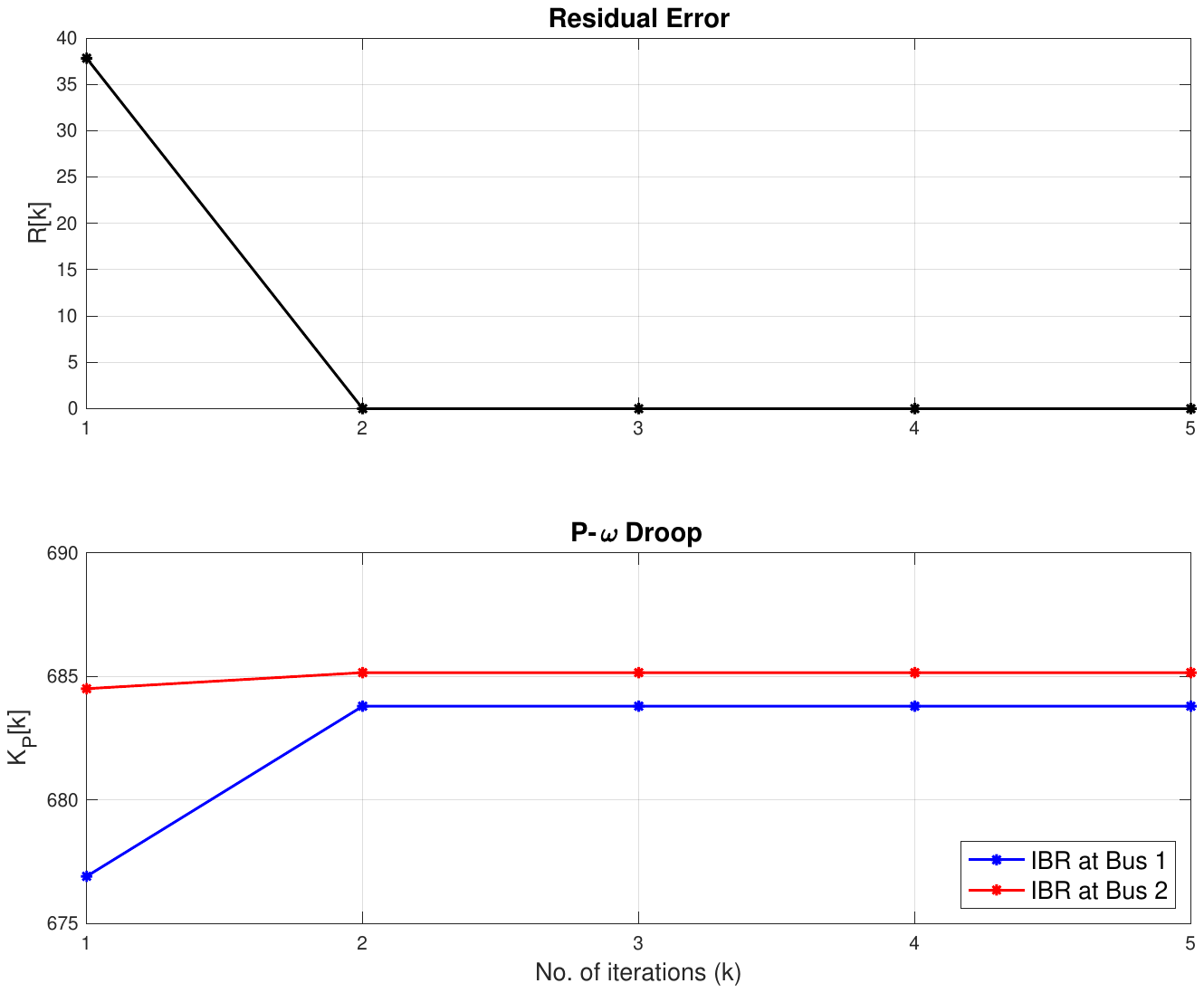}}
    \caption{Performance of the iterative algorithm (Low-impedance).}
    \label{fig:alg2} 
\end{figure}

Performance of the iterative algorithm is illustrated in Fig. \ref{fig:alg2}. Convergence is achieved in two steps and the resulting optimal gains are $\mathbf{K_P} = [683.795, 685.155]$. 
Notably, these gains are higher than those obtained in the base impedance scenario, consistent with the fact that increasing the P-$\omega$ droop gain in the unified inverter improves system stability under stronger grid conditions \cite{geng2022unified}.

\begin{figure}[ht!]
    \hspace{0ex} \centerline{\includegraphics[scale=0.37]{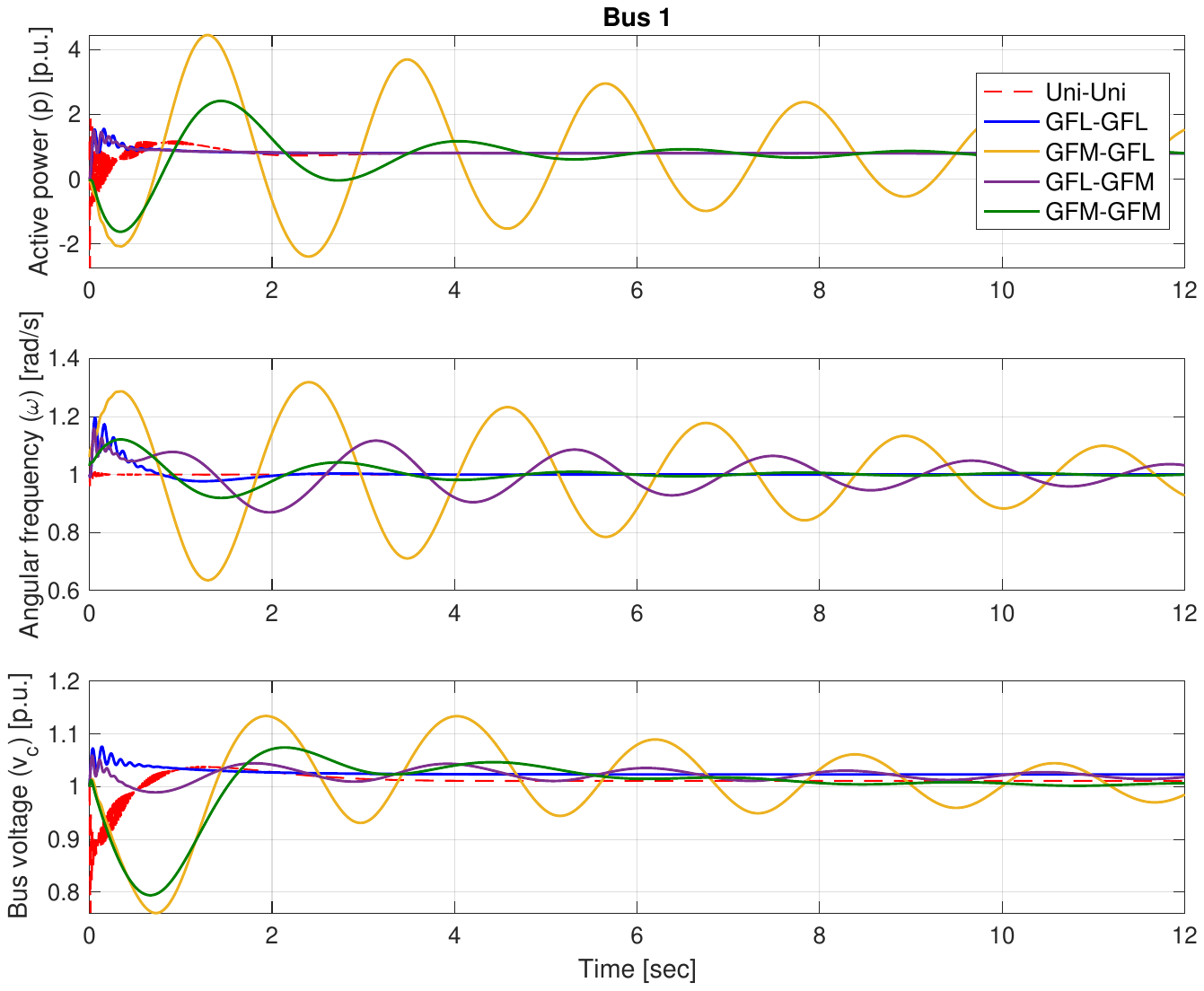}}
    \caption{Trajectories of active power, angular frequency, and voltage at Bus 1 (Low-impedance).}
    \label{fig:cb1} 
\end{figure}

\begin{figure}[ht!]
    \hspace{0ex} \centerline{\includegraphics[scale=0.37]{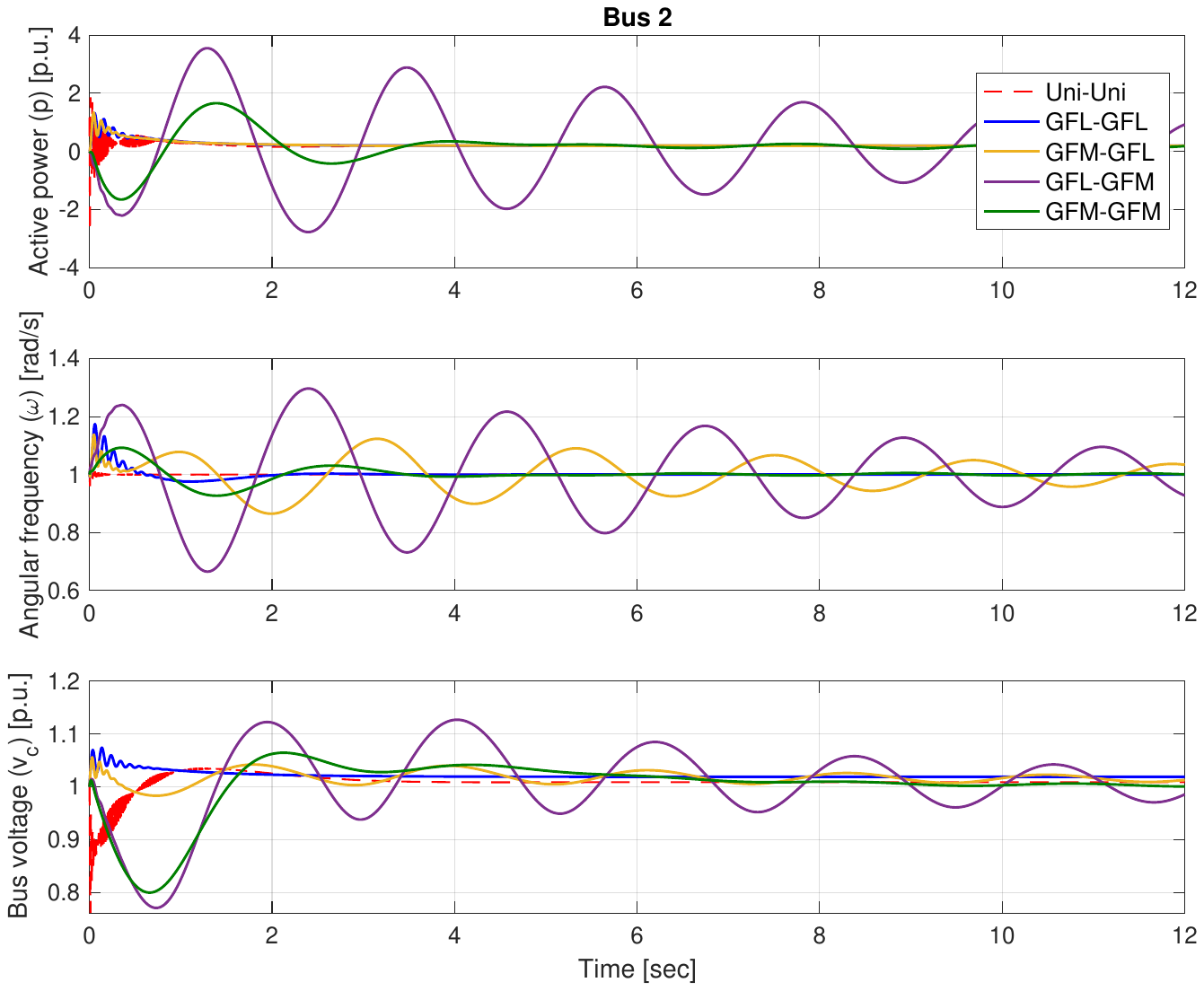}}
    \caption{Trajectories of active power, angular frequency, and voltage at Bus 2 (Low-impedance).}
    \label{fig:cb2} 
\end{figure}

The simulation results are shown in Figs. \ref{fig:cb1} and \ref{fig:cb2}, corresponding to buses 1 and 2, respectively. As illustrated in Fig. \ref{fig:cb1}, the GFL-only system outperforms other configurations across all metrics, consistent with expectations in a strong grid scenario. Here, the optimal P-$\omega$ droop determined for the unified-only system is higher than that of the base impedance case, which inversely corresponds to a lower $\omega$-P droop for GFM inverters. Consequently, the allocation strategy should favor positioning GFL inverters in these locations rather than GFM ones to achieve better performance.

\subsubsection{Case 3: High-impedance network}
In a weaker grid scenario, the line impedances are increased {\color{black} by $27\%$, $23\%$ and $23\%$ (relative to the base values) for lines 1-2, 1-3 and 2-3, respectively,} as given in Table~\ref{tab:lin3}. 
\begin{table}[ht!]
  \setlength\tabcolsep{5pt}
  \setlength\extrarowheight{2pt}
  \centering
  \caption{Transmission line data for 3-bus network (in per units)}
    \begin{tabular}{ccc}
      \Xhline{2\arrayrulewidth} 
      $y_{12}$ & $y_{13}$ & $y_{23}$ \\
      \hline
      0.0736 - $\iu$2.3787 & 2.9626 - $\iu$10.2552 & 2.9626 - $\iu$10.2552 \\
      \Xhline{2\arrayrulewidth} 
    \end{tabular}
    \label{tab:lin3}
\end{table}

\begin{figure}[ht!]
    \hspace{0ex} \centerline{\includegraphics[scale=0.37]{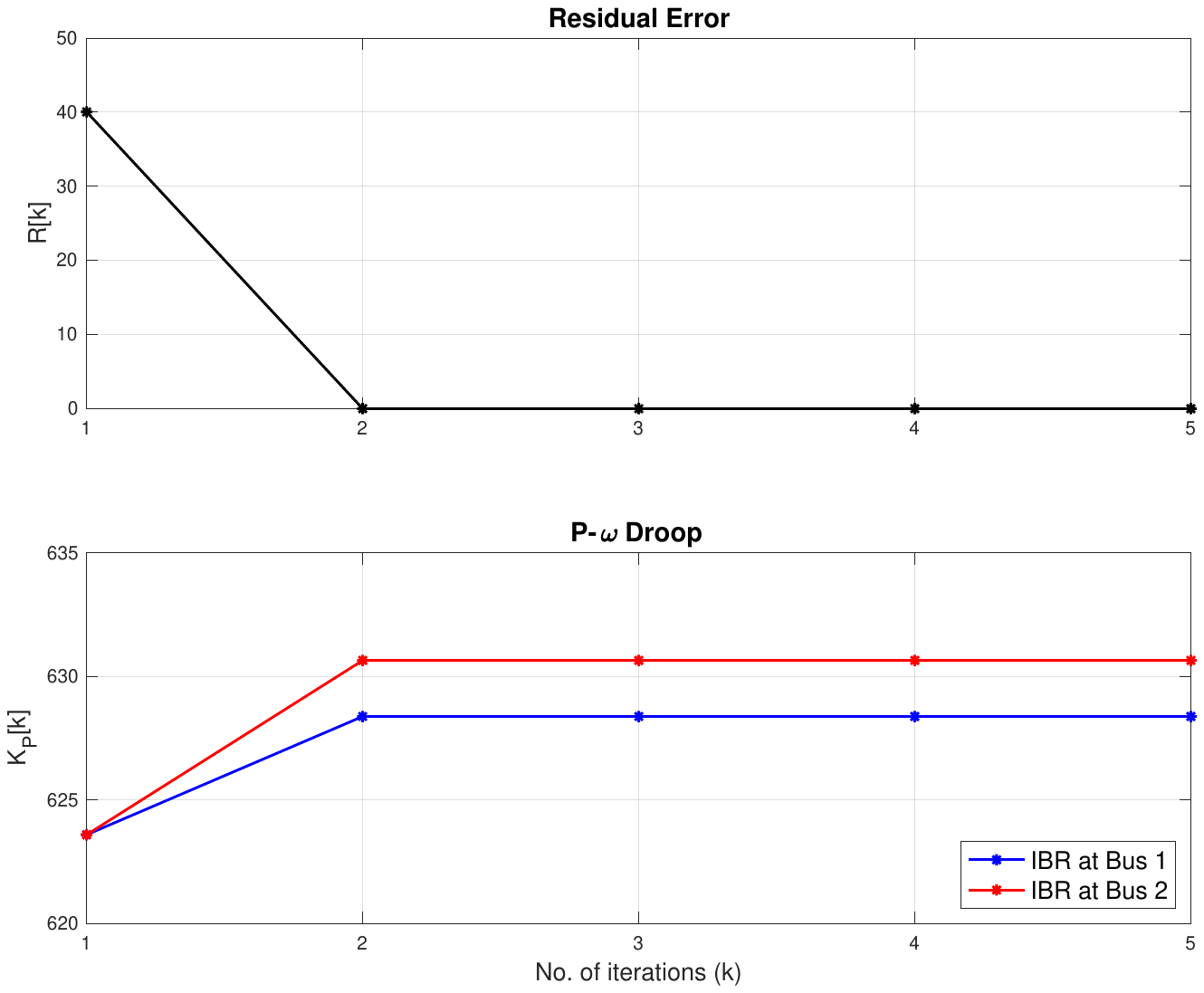}}
    \caption{Performance of the iterative algorithm (High-impedance).}
    \label{fig:alg3} 
\end{figure}

Performance of the iterative algorithm is given in Fig. \ref{fig:alg3}. Convergence is achieved in two steps and the optimal droop gains are $\mathbf{K_P} = [628.380, 630.649]$. In this scenario, the optimal P-$\omega$ droop for the unified-only system is lower than that of the base impedance network, which corresponds to a higher $\omega$-P droop for GFM inverters. Consequently, the allocation would prefer GFM resources to improve system stability. 

\begin{figure}[ht!]
    \hspace{0ex} \centerline{\includegraphics[scale=0.37]{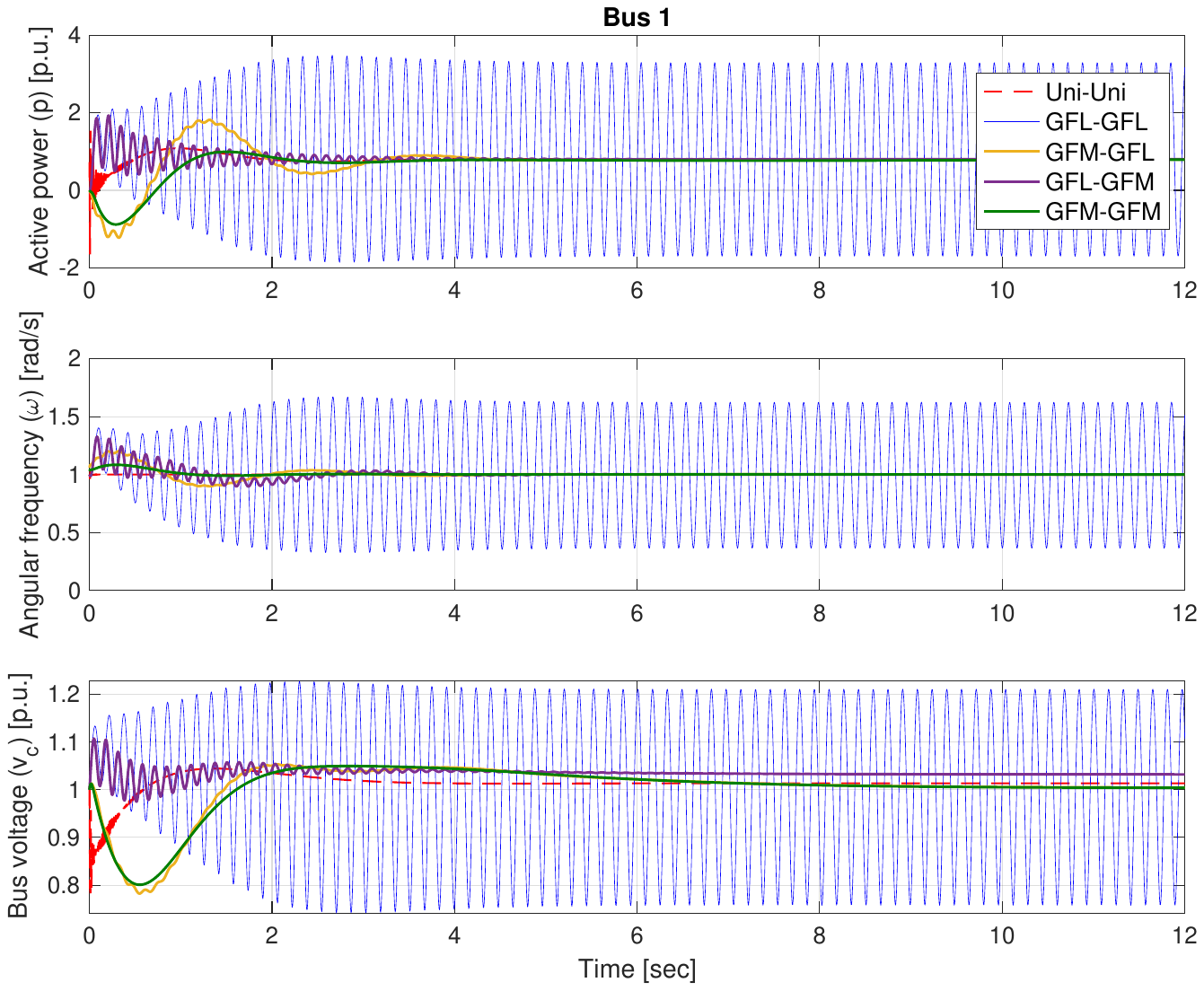}}
    \caption{Trajectories of active power, angular frequency, and voltage at Bus 1 (High-impedance).}
    \label{fig:cc1} 
\end{figure}

\begin{figure}[ht!]
    \hspace{0ex} \centerline{\includegraphics[scale=0.37]{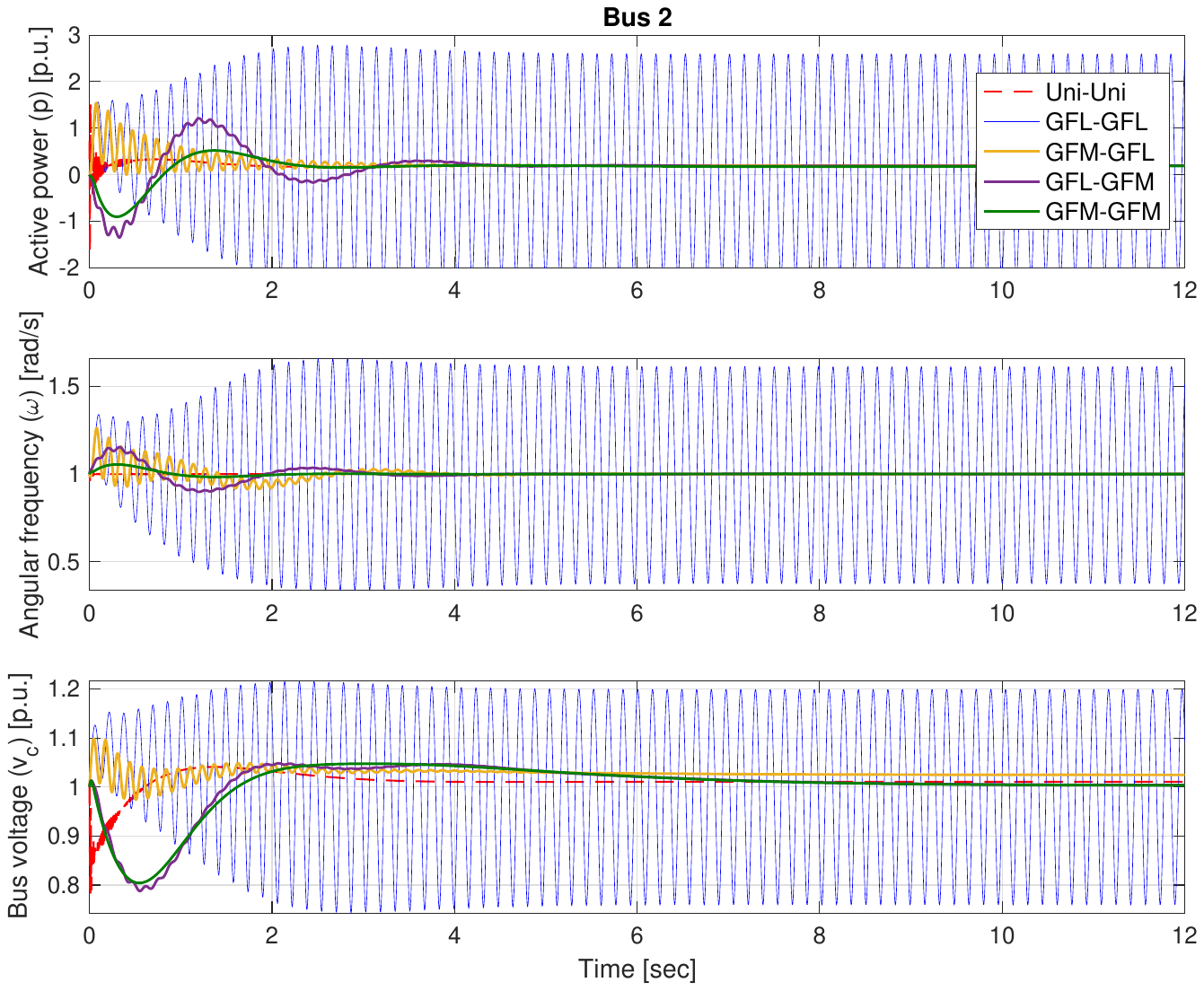}}
    \caption{Trajectories of active power, angular frequency, and voltage at Bus 2 (High-impedance).}
    \label{fig:cc2} 
\end{figure}

The simulation results are presented in Figs. \ref{fig:cc1} and \ref{fig:cc2}, for buses 1 and 2, respectively. As shown in Fig. \ref{fig:cc1}, the GFM-only system consistently gives superior performance across all metrics, aligning with expectations for a weak grid scenario, while the GFL-only system shows the poorest performance. Although the mixed GFM-GFL configuration achieves a settling time similar to the GFM-only system, the oscillations are much more pronounced.

\section{Conclusion}\label{sec:conclusion}
This paper leveraged a novel unified inverter control to devise a systematic and efficient way to allocate grid-forming (GFM) and grid-following (GFL) inverter-based resources (IBRs) in power networks. 
This method was motivated by the fact that the unified controller inherits the characteristics of both GFM and GFL inverters. A single parameter, the P-$\omega$ droop of each unified inverter was identified as the decision variable. 
An optimization problem was subsequently formulated to ensure system stability and maximal energy dissipation during dynamic transients. 

An iterative algorithm was proposed to solve the optimization problem. The optimal droop gains guided the allocation of GFM and GFL IBRs in the network. Numerical results showed that the proposed method was effective in identifying the optimal IBR placement for various network conditions. Specifically, as demonstrated in a three-bus network, GFM inverters are preferable in weak grid conditions to enhance stability, while GFL inverters are preferable in strong grid conditions.  
In the future, we will establish theoretical convergence guarantees for the proposed algorithm and investigate larger power systems.

\section*{Acknowledgement}
The authors acknowledge the valuable discussions with Professor Ian Hiskens about the ideas presented in this paper.

\bibliographystyle{ieeetr}
\bibliography{Bibliography.bib}
\balance

\section*{Appendix}
\subsection{Unified Grid-Forming/Following Inverter} \label{app:detailed_model_unified}
\subsubsection{Nomenclature}
The variables and parameters used in the differential and algebraic equations are indexed below.\\
\texttt{Dynamic states (12):}\\
\begin{small}
$x_1\ :\ \tilde{p} \rightarrow$ Filtered active power\\
$x_2\ :\ \tilde{q} \rightarrow$ Filtered reactive power\\
$x_3\ :\ \phi_d \rightarrow$ d-axis voltage controller state\\
$x_4\ :\ \eta \rightarrow$ Active power controller state\\
$x_5\ :\ \delta \rightarrow$ Angle between $v_t$ and $v_c$\\
$x_6\ :\ \zeta \rightarrow$ PLL integrator state\\
$x_7\ :\ \theta_{PLL} \rightarrow$ PLL phase angle\\
$x_8\ :\ \gamma_d \rightarrow$ d-axis current controller state\\
$x_9\ :\ i_{t,d} \rightarrow$ (Local) d-axis IBR output current $i_t$\\
$x_{10}\ :\ i_{t,d} \rightarrow$ (Local) q-axis IBR output current $i_t$\\
$x_{11}\ :\ v_{c,d} \rightarrow$ (Local)  d-axis filter output voltage $v_c$\\
$x_{12}\ :\ v_{c,q} \rightarrow$ (Local) q-axis filter output voltage $v_c$
\end{small}
\newline\newline
\texttt{Algebraic states (20):}\\
\begin{small}
$y_1\ :\ \omega \rightarrow$ Angular frequency\\
$y_2\ :\ \theta_c \rightarrow$ Angle of $v_c$\\
$y_3\ :\ \theta_t \rightarrow$ Angle of $v_t$\\
$y_4\ :\ p_0 \rightarrow$ Droop-modified active power setpoint\\
$y_5\ :\ \omega_{PLL} \rightarrow$ PLL angular frequency\\
$y_6\ :\ v^{*}_{c,d} \rightarrow$ d-axis voltage setpoint $v^{*}_c$\\
$y_7\ :\ i_{t,d}^{*} \rightarrow$ d-axis current setpoint $i_t^{*}$\\
$y_8\ :\ i_{g,d} \rightarrow$ d-axis filter output current $i_g$\\
$y_9\ :\ i_{g,q} \rightarrow$ q-axis filter output current $i_g$\\
$y_{10}\ :\ i_{g,D} \rightarrow$ D-axis filter output current $i_g$\\
$y_{11}\ :\ i_{g,Q} \rightarrow$ Q-axis filter output current $i_g$\\
$y_{12}\ :\ v_{t,d} \rightarrow$ d-axis IBR output voltage $v_t$\\
$y_{13}\ :\ v_{t,q} \rightarrow$ q-axis IBR output voltage $v_t$\\
$y_{14}\ :\ v_t \rightarrow$ IBR output voltage\\
$y_{15}\ :\ v_{c,D} \rightarrow$ D-axis filter output voltage $v_c$\\
$y_{16}\ :\ v_{c,Q} \rightarrow$ Q-axis filter output voltage $v_c$\\
$y_{17}\ :\ p \rightarrow$ Active power output from filter\\
$y_{18}\ :\ q \rightarrow$ Reactive power output from filter\\
$y_{19}\ :\ u_{dc} \rightarrow$ DC-link voltage\\
$y_{20}\ :\ i_{dc} \rightarrow$ DC-link current
\end{small}
\newline\newline
\texttt{Parameters (21):}\\
\begin{small}
$p_1\ :\ p^{*} \rightarrow$ Active power setpoint \\
$p_2\ :\ q^{*} \rightarrow$ Reactive power setpoint \\
$p_3\ :\ \omega_0 \rightarrow$ Angular frequency setpoint of IBR \\
$p_4\ :\ V_0 \rightarrow$ Voltage setpoint of IBR \\
$p_5\ :\ K_P \rightarrow$ Active (P-$\omega$) droop coefficient \\
$p_6\ :\ K_Q \rightarrow$ Reactive (V-Q) droop coefficient \\
$p_7\ :\ \omega_{pc} \rightarrow$  Active power filter 3dB cut-off frequency \\
$p_8\ :\ \omega_{qc} \rightarrow$  Reactive power filter 3dB cut-off frequency \\
$p_9\ :\ \omega_b \rightarrow$ Base angular frequency \\
$p_{10}\ :\ K^P_{VC} \rightarrow$  Proportional gain of voltage controller \\
$p_{11}\ :\ K^I_{VC} \rightarrow$  Integral gain of voltage controller \\
$p_{12}\ :\ K^F_{VC} \rightarrow$  Feed-forward gain of voltage controller \\
$p_{13}\ :\ K^P_{CC} \rightarrow$  Proportional gain of current controller \\
$p_{14}\ :\ K^I_{CC} \rightarrow$  Integral gain of current controller \\
$p_{15}\ :\ K^F_{CC} \rightarrow$  Feed-forward gain of current controller \\
$p_{16}\ :\ K_{PC}^P \rightarrow$ Proportional gain of active power controller \\
$p_{17}\ :\ K_{PC}^I \rightarrow$ Integral gain of active power controller \\
$p_{18}\ :\ K_{PLL}^P \rightarrow$ Proportional gain of PLL \\
$p_{19}\ :\ K_{PLL}^I \rightarrow$ Integral gain of PLL \\
$p_{20}\ :\ C_f \rightarrow$ Filter capacitance \\
$p_{21}\ :\ L_f \rightarrow$ Filter inductance
\end{small}

\subsubsection{System Description}
The differential-algebraic equations (DAEs) outlined below describe the system's behaviour.
\begin{figure*}[ht!]
    \centerline{\includegraphics[scale=0.52]{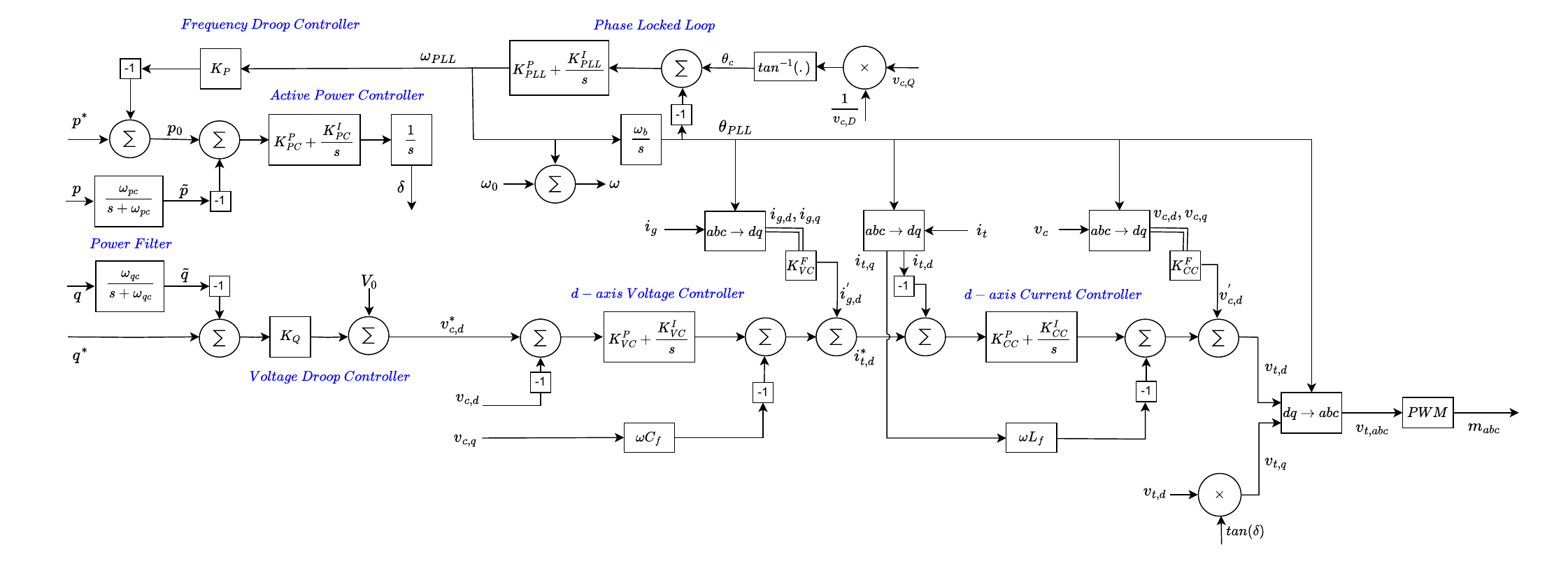}}
    \caption{Detailed control architecture of the unified inverter.}
    \label{fig:control_uni} 
\end{figure*} \\ \noindent
\texttt {Differential equations:}
\begin{flalign}
    \dot{\tilde{p}} &\ =\ -\omega_{pc} \tilde{p} + p \omega_{pc} \label{eqn:su1}&&
    \\
    \dot{\tilde{q}} &\ =\ -\omega_{qc} \tilde{q} + q \omega_{qc} \label{eqn:su2}
    \\
    \dot{\phi_d} &\ =\ v^{*}_{c,d} - v_{c,d} \label{eqn:su3}
    \\
    \dot{\eta} &\ =\ p_0 - \tilde{p} \label{eqn:su4}
    \\
    \dot{\delta} &\ =\ K^P_{PC}(p_0 - \tilde{p}) + K^I_{PC}\eta \label{eqn:su5}
    \\
    \dot{\zeta} &\ =\ \theta_c - \theta_{PLL} \label{eqn:su6}
    \\
    \dot{\theta_{PLL}} &\ =\ \omega_{PLL} \omega_b \label{eqn:su7}
    \\
    \dot{\gamma_d} &\ =\ i^{*}_{t,d} - i_{t,d} \label{eqn:su8}
    \\
    \dot{i_{t,d}} &\ =\ \frac{\omega_b}{L_f}(v_{t,d} - v_{c,d}) + \omega_b \omega i_{t,q} \label{eqn:su9}
    \\
    \dot{i_{t,q}} &\ =\ \frac{\omega_b}{L_f}(v_{t,q} - v_{c,q}) - \omega_b \omega i_{t,d} \label{eqn:su10}
    \\
    \dot{v_{c,d}} &\ =\ \frac{\omega_b}{C_f}(i_{t,d} - i_{g,d}) + \omega_b \omega v_{c,q} \label{eqn:su11}
    \\
    \dot{v_{c,q}} &\ =\ \frac{\omega_b}{C_f}(i_{t,q} - i_{g,q}) - \omega_b \omega v_{c,d} \label{eqn:su12}
\end{flalign}

\noindent
\texttt{Algebraic equations:}
\begin{align}
    &\ \omega - \omega_0 - \omega_{PLL} = 0 \label{eqn:au1}
    \\&\
    p_0 - p^{*} - K_P\omega_{PLL} = 0\label{eqn:au2}
    \\&\
    v^{*}_{c,d} - V_0 - K_Q(q^{*} - \tilde{q}) = 0 \label{eqn:au3}
    \\&\
    \delta - \theta_t + \theta_{PLL} = 0 \label{eqn:au4}
    \\&\
    \omega_{PLL} - K^P_{PLL}(\theta_c - \theta_{PLL}) + K^I_{PLL}\zeta = 0 \label{eqn:au5}
    \\&\
    p - v_{c,d}i_{g,d} - v_{c,q}i_{g,q} = 0 \label{eqn:au6}
    \\&\
    q - v_{c,q}i_{g,d} + v_{c,d}i_{g,q} = 0 \label{eqn:au7}
    \\&\
    v_{c,D}tan(\theta_c) - v_{c,Q} = 0 \label{eqn:au8}
    \\&\
    v_{c,D} - v_{c,d}\cos(\theta_{PLL}) + v_{c,q}\sin(\theta_{PLL}) = 0 \label{eqn:au9}
    \\&\ 
    v_{c,Q} - v_{c,d}\sin(\theta_{PLL}) - v_{c,q}\cos(\theta_{PLL}) = 0 \label{eqn:au10} 
    \\&\ 
    i_{g,D} - i_{g,d}\cos(\theta_{PLL}) + i_{g,q}\sin(\theta_{PLL}) = 0 \label{eqn:au11}
    \\&\
    i_{g,Q} - i_{g,d}\sin(\theta_{PLL}) - i_{g,q}\cos(\theta_{PLL}) = 0 \label{eqn:au12} 
    \\&\
    i^{*}_{t,d} \!-\! K^P_{VC}(v^{*}_{c,d} \!-\! v_{c,d}) \!+\! K^I_{VC}\phi_d \!+\! K^F_{VC}i_{g,d} \!-\! \omega C_f v_{c,q} = 0 \label{eqn:au13}
    \\&\
    v_{t,d} \!-\! K^P_{CC}(i^{*}_{t,d} \!-\! i_{t,d}) \!+\! K^I_{CC}\gamma_d \!+\! K^F_{CC}v_{c,d} \!-\! \omega L_f i_{t,q} = 0 \label{eqn:au14}
    \\&\
    v_t \cos(\theta_t) - v_{t,d} \cos(\theta_{PLL}) + v_{t,q}\sin(\theta_{PLL}) = 0 \label{eqn:au15}
    \\&\
    v_t \sin(\theta_t) - v_{t,d} \sin(\theta_{PLL}) - v_{t,q}\cos(\theta_{PLL}) = 0 \label{eqn:au16}
    \\&\
    u_{dc}i_{dc} - v_{t,d}i_{t,d} - v_{t,q}i_{t,q} = 0 \label{eqn:au17}
\end{align}

\subsection{Grid Following (GFL) Inverter}\label{app:detailed_model_gfl}
\begin{figure}[ht!]
    \centerline{\includegraphics[scale=0.8]{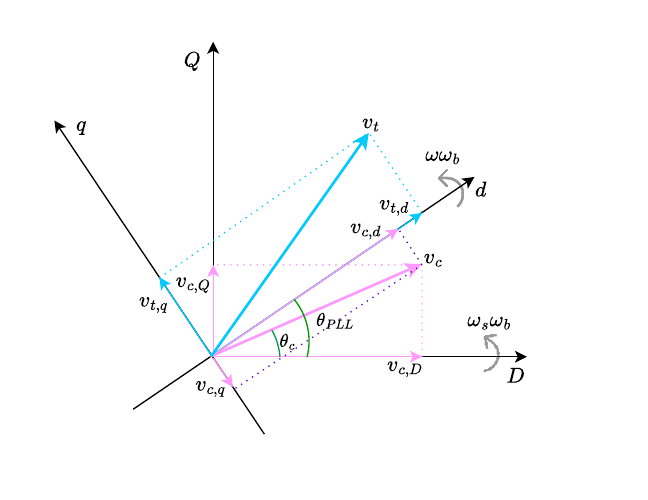}}
    \caption{Phasor diagram and reference frames of the GFL inverter.}
    \label{fig:gfl_ph} 
\end{figure}

\begin{figure*}[ht!]
    \centerline{\includegraphics[scale=0.68]{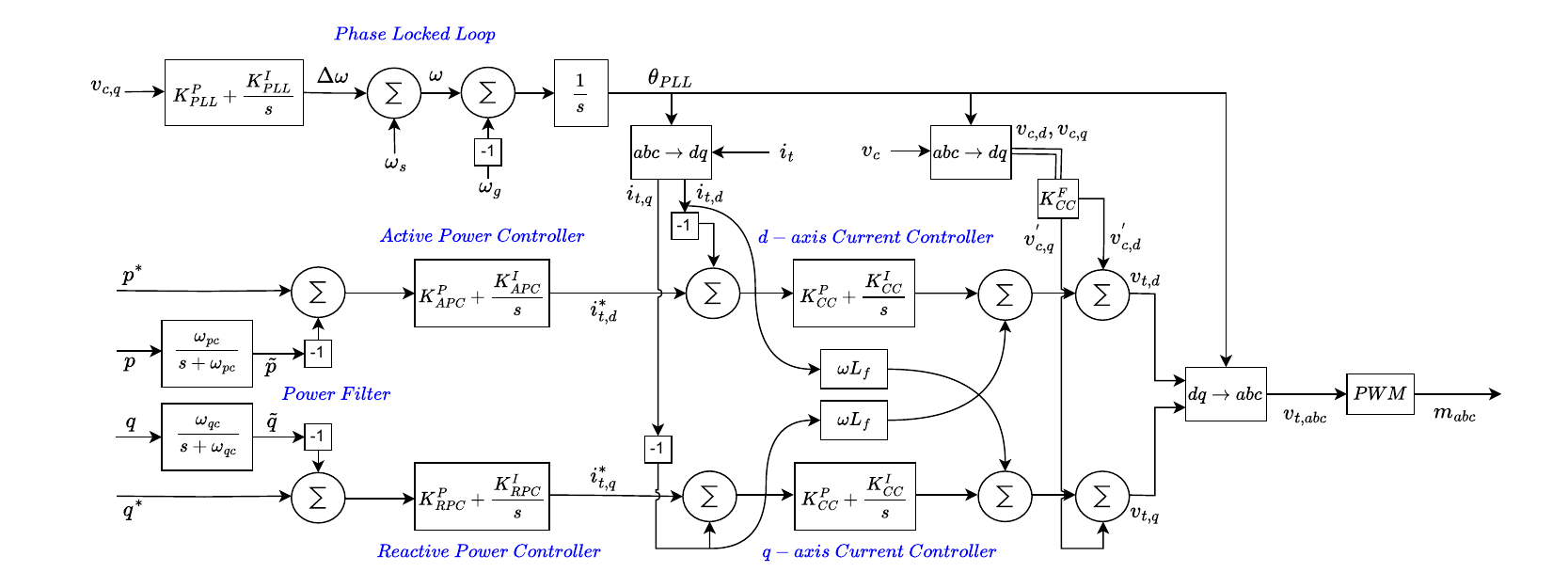}}
    \caption{Detailed control architecture of the GFL inverter.}
    \label{fig:control_gfl} 
\end{figure*}

The GFL inverter consists of a VSC with an output LC filter, as described by equations \eqref{eqn:sl1}-\eqref{eqn:sl2} and \eqref{eqn:sl5}-\eqref{eqn:sl6}. Its control system employs a hierarchical structure with an outer power control loop and an inner current control loop, illustrated in Fig. \ref{fig:control_gfl}. The outer loop includes active (APC) \eqref{eqn:sl9} and reactive (RPC) \eqref{eqn:sl10} power controllers, which use power measurements from \eqref{eqn:al7}-\eqref{eqn:al8}. The reference signals generated by the outer loops \eqref{eqn:al12}-\eqref{eqn:al13} are fed into the inner current-control (CC) loops, described by \eqref{eqn:sl3}-\eqref{eqn:sl4} and \eqref{eqn:al3}-\eqref{eqn:al4}. The DC power balance is governed by \eqref{eqn:al14}. All control loops are implemented in a local dq-reference frame. A phase-locked loop (PLL) \eqref{eqn:sl7} generates the angle \eqref{eqn:sl8} and frequency \eqref{eqn:al9}, which are used to convert three-phase voltage and current signals into dq quantities. The transformation from local to global DQ-frame is detailed in equations \eqref{eqn:al1}-\eqref{eqn:al2}, \eqref{eqn:al5}-\eqref{eqn:al6}, and \eqref{eqn:al10}-\eqref{eqn:al11}, and illustrated in Fig. \ref{fig:gfl_ph}. The differential and algebraic equations governing the system in Fig. \ref{fig:setup_gfl} along with the nomenclature are outlined below.
Table \ref{tab:tgfl} presents the GFL parameters used in the simulations, adapted from \cite{pal2023large} with additional modifications. Unless specified otherwise, all parameters are expressed in per unit.
\begin{table}[ht!]
    \centering
  \setlength\tabcolsep{2.3pt}
  \setlength\extrarowheight{2pt}
  \caption{Grid-following (GFL) inverter parameters}
    \begin{tabular}{cccccccc}
      \Xhline{2\arrayrulewidth} 
      $\omega_s$ & $R_f$ & $L_f$ & $C_f$ & $K^P_{APC}$ &  $K^I_{APC}$ & $K^P_{RPC}$ & $K^I_{RPC}$ \vspace{0.8ex}\\
      \hline
      1 & 0.0072 & 0.2724 & 0.2612 & 1 & 3 & -3 & -4 \\
      \hhline{|========|}
       $K^P_{CC}$ & $K^I_{CC}$ & $K^P_{PLL}$ & $K^I_{PLL}$ & $\omega_g$ & $\omega_{pc}$ & $\omega_{qc}$ & \vspace{0.8ex}\\ 
       \hline
      1.443 & 14.43 & 2.65 & 6.5 & 1 & 332.8 rad/s & 732.8 rad/s & \\ 
      \Xhline{2\arrayrulewidth} 
    \end{tabular}
    \label{tab:tgfl}
\end{table}

\subsubsection{Nomenclature}
The variables and parameters used in the differential and algebraic equations are indexed below.\\
\texttt{Dynamic states (12):}\\
\begin{small}
$x_1\ :\ i_{t,d} \rightarrow$ (Local) d-axis IBR output current $i_t$\\
$x_2\ :\ i_{t,q} \rightarrow$ (Local) q-axis IBR output current $i_t$\\
$x_3\ :\ \gamma_d \rightarrow$ d-axis current controller state\\
$x_4\ :\ \gamma_q \rightarrow$  q-axis current controller state\\
$x_5\ :\ v_{c,D} \rightarrow$ (Global) D-axis filter output voltage $v_c$\\
$x_6\ :\ v_{c,Q} \rightarrow$ (Global) Q-axis filter output voltage $v_c$\\
$x_7\ :\ \gamma_{PLL} \rightarrow$ PLL integrator state\\
$x_8\ :\ \theta_{PLL} \rightarrow$ PLL phase angle\\
$x_9\ :\ \phi_d \rightarrow$ d-axis (active) power controller state\\
$x_{10}\ :\ \phi_q \rightarrow$ q-axis (reactive) power controller state\\
$x_{11}\ :\ \tilde{p} \rightarrow$ Filtered active power\\
$x_{12}\ :\ \tilde{q} \rightarrow$ Filtered reactive power
\end{small}
\newline\newline
\texttt{Algebraic states (17):}\\
\begin{small}
$y_1\ :\ v_{t,d} \rightarrow$ d-axis IBR output voltage $v_t$\\
$y_2\ :\ v_{t,q} \rightarrow$ q-axis IBR output voltage $v_t$\\
$y_3\ :\ \omega \rightarrow$ IBR angular frequency determined by PLL\\
$y_4\ :\ i^{*}_{t,d} \rightarrow$ d-axis current setpoint $i^{*}_t$\\
$y_5\ :\ i^{*}_{t,q} \rightarrow$  q-axis current setpoint $i^{*}_t$\\
$y_6\ :\ i_{t,D} \rightarrow$ D-axis IBR output current $i_t$\\
$y_7\ :\ i_{t,Q} \rightarrow$ Q-axis IBR output current $i_t$\\
$y_8\ :\ v_{c,d} \rightarrow$ d-axis filter output voltage $v_c$\\
$y_9\ :\ v_{c,q} \rightarrow$ q-axis filter output voltage $v_c$\\
$y_{10}\ :\ p \rightarrow$ Active power output from filter\\
$y_{11}\ :\ q \rightarrow$ Reactive power output from filter\\
$y_{12}\ :\ i_{g,d} \rightarrow$ d-axis filter output current $i_g$\\
$y_{13}\ :\ i_{g,q} \rightarrow$ q-axis filter output current $i_g$\\
$y_{14}\ :\ i_{g,D} \rightarrow$ D-axis filter output current $i_g$\\
$y_{15}\ :\ i_{g,Q} \rightarrow$ Q-axis filter output current $i_g$\\
$y_{16}\ :\ u_{dc} \rightarrow$ DC-link voltage\\
$y_{17}\ :\ i_{dc} \rightarrow$ DC-link current
\end{small}
\newline\newline
\texttt{Parameters (17):}\\
\begin{small}
$p_1\ :\ L_f \rightarrow$ Filter inductance \\
$p_2\ :\ C_f \rightarrow$ Filter capacitance \\
$p_3\ :\ K_{PLL}^P \rightarrow$ Proportional gain of PLL \\
$p_4\ :\ K_{PLL}^I \rightarrow$ Integral gain of PLL \\
$p_5\ :\ \omega_s \rightarrow$ Synchronous angular frequency for IBR \\
$p_6\ :\ K_{CC}^P \rightarrow$  Proportional gain of current controller \\
$p_7\ :\ K_{CC}^I \rightarrow$  Integral gain of current controller \\
$p_8\ :\ K_{CC}^F \rightarrow$  Feed-forward gain of current controller \\
$p_9\ :\ K_{APC}^P \rightarrow$  Proportional gain of active power controller \\
$p_{10}\ :\ K_{APC}^I \rightarrow$  Integral gain of active power controller \\
$p_{11}\ :\ K_{RPC}^P \rightarrow$ Proportional gain of reactive power controller \\
$p_{12}\ :\ K_{RPC}^I \rightarrow$ Integral gain of reactive power controller \\
$p_{13}\ :\ \omega_g \rightarrow$ Grid angular frequency \\
$p_{14}\ :\ p^{*} \rightarrow$ Active power setpoint \\
$p_{15}\ :\ q^{*} \rightarrow$ Reactive power setpoint \\
$p_{16}\ :\ \omega_{pc} \rightarrow$ Active power filter 3dB cut-off frequency \\
$p_{17}\ :\ \omega_{qc} \rightarrow$ Reactive power filter 3dB cut-off frequency 
\end{small}

\subsubsection{System Description}
The differential-algebraic equations (DAEs) outlined below describe the system's behavior.\\ \noindent
\texttt{Differential equations:}
\begin{flalign}
        \dot{i_{t,d}} &\ =\ \omega i_{t,q} + \frac{1}{L_f}(v_{t,d} - v_{c,d}) \label{eqn:sl1} &&
        \\
        \dot{i_{t,q}} &\ =\ - \omega i_{t,d} + \frac{1}{L_f}(v_{t,q} - v_{c,q}) \label{eqn:sl2}
        \\
        \dot{\gamma_d} &\ =\ i^{*}_{t,d} - i_{t,d} \label{eqn:sl3}
        \\
        \dot{\gamma_q} &\ =\ i^{*}_{t,q} - i_{t,q} \label{eqn:sl4}
        \\
        \dot{v}_{c,D} &\ =\ \omega_g v_{c,Q} + \frac{1}{C_f}(i_{t,D} - i_{g,D}) \label{eqn:sl5}
        \\
        \dot{v}_{c,Q} &\ =\ -\omega_g v_{c,D} + \frac{1}{C_f}(i_{t,Q} - i_{g,Q}) \label{eqn:sl6}
        \\
        \dot{\gamma}_{PLL} &\ =\ v_{c,q} \label{eqn:sl7}
        \\
        \dot{\theta_{PLL}} &\ =\ K^P_{PLL}v_{c,q} + K^I_{PLL}\gamma_{PLL} + \omega_s - \omega_g \label{eqn:sl8}
        \\
        \dot{\phi}_d &\ =\ p^{*} - p \label{eqn:sl9}
        \\
        \dot{\phi}_q &\ =\ q^{*} - q \label{eqn:sl10}
        \\
        \dot{\tilde{p}} &\ =\ -\omega_{pc} \tilde{p} + p \omega_{pc} \label{eqn:sl11}
        \\
        \dot{\tilde{q}} &\ =\ -\omega_{qc} \tilde{q} + q \omega_{qc} \label{eqn:sl12}
\end{flalign}
\texttt{Algebraic equations:}
\begin{align}
    &\ v_{c,d} - v_{c,D}\cos(\theta_{PLL}) - v_{c,Q}\sin(\theta_{PLL}) = 0 \label{eqn:al1}
        \\&\
        v_{c,q} + v_{c,D}\sin(\theta_{PLL}) - v_{c,Q}\cos(\theta_{PLL}) = 0 \label{eqn:al2}
        \\&\
        v_{t,d} - K^P_{CC}(i^{*}_{t,d}\!-\!i_{t,d}) - K^I_{CC} \gamma_d + \omega L_f i_{t,q} - K^F_{CC}v_{c,d} = 0 \label{eqn:al3}
        \\&\
        v_{t,q} - K^P_{CC}(i^{*}_{t,q}\!-\!i_{t,q}) - K^I_{CC} \gamma_q - \omega L_f i_{t,d} - K^F_{CC}v_{c,q} = 0 \label{eqn:al4}
        \\&\
         i_{t,D} - i_{t,d}\cos(\theta_{PLL}) + i_{t,q}\sin(\theta_{PLL}) = 0 \label{eqn:al5}
        \\&\
        i_{t,Q} - i_{t,d}\sin(\theta_{PLL}) - i_{t,q}\cos(\theta_{PLL}) = 0 \label{eqn:al6}
        \\&\ p - v_{c,d}i_{g,d} - v_{c,q}i_{g,q} = 0 \label{eqn:al7}\\
    &\ q - v_{c,q}i_{g,d} + v_{c,d}i_{g,q} = 0 \label{eqn:al8}
        \\&\ \omega - \omega_s - K^P_{PLL}v_{c,q} - K^I_{PLL}\gamma_{PLL} = 0 \label{eqn:al9}
        \\&\
        i_{g,d} - i_{g,D}\cos(\theta_{PLL}) - i_{g,Q}\sin(\theta_{PLL}) = 0 \label{eqn:al10}
        \\&\
        i_{g,q} + i_{g,D}\sin(\theta_{PLL}) - i_{g,Q}\cos(\theta_{PLL}) = 0 \label{eqn:al11}
        \\&\
        i_{t,d}^{*} - K^P_{APC}(p^{*}-p) - K^I_{APC}\phi_d = 0 \label{eqn:al12}
        \\&\
        i_{t,q}^{*} - K^P_{RPC}(q^{*}-q) - K^I_{RPC}\phi_q = 0 \label{eqn:al13}
        \\&\ 
        u_{dc}i_{dc} - v_{t,d}i_{t,d} - v_{t,q}i_{t,q} = 0 \label{eqn:al14}
\end{align}

\subsection{Grid Forming (GFM) Inverter}\label{app:detailed_model_gfm}
\begin{figure*}[ht!]
    \centerline{\includegraphics[scale=0.52]{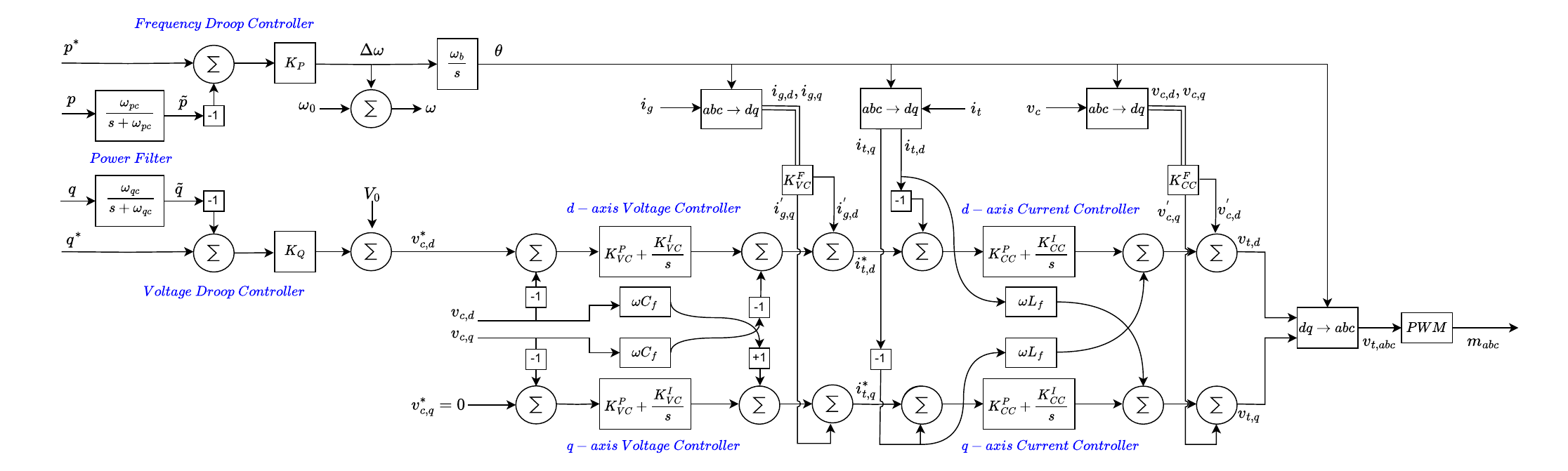}}
    \caption{Detailed control architecture of the GFM inverter.}
    \label{fig:control_gfm} 
\end{figure*}

The general structure of a three-phase GFM inverter system is illustrated in Fig. \ref{fig:setup_gfm}. This setup includes a VSC with an output LC filter, characterized by equations \eqref{eqn:sm8}-\eqref{eqn:sm11}.
The controller features a nested structure consisting of an outer voltage control loop and an inner current control loop, as shown in Fig. \ref{fig:control_gfm}.  

\begin{figure}[ht!]
    \hspace{-4ex} \centerline{\includegraphics[scale=0.8]{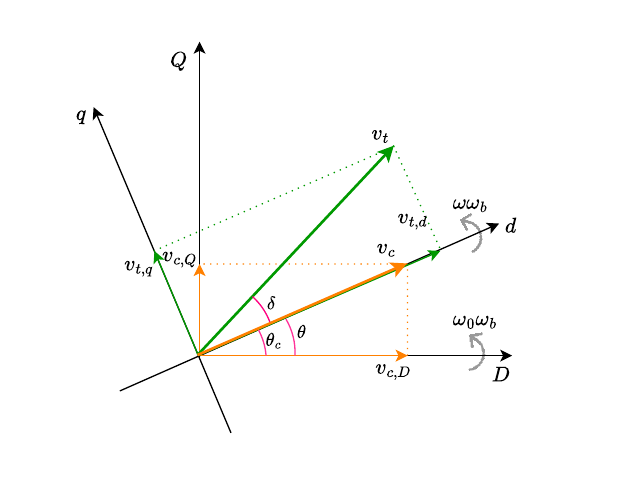}}
    \caption{Phasor diagram and reference frames of the GFM inverter.}
    \label{fig:gfm_ph} 
\end{figure}

Each control loop operates in a direct-quadrature (dq) reference frame, utilizing Park’s transformation, as described in \eqref{eqn:am7}-\eqref{eqn:am12}, and illustrated in Fig. \ref{fig:gfm_ph}. 
The outer loop includes voltage controllers (VC), governed by \eqref{eqn:sm4}-\eqref{eqn:sm5}. Reference signals generated by the outer loop, as defined in \eqref{eqn:am3}-\eqref{eqn:am4}, are passed to the inner current control (CC) loops, which are described by \eqref{eqn:sm6}-\eqref{eqn:sm7} and \eqref{eqn:am5}-\eqref{eqn:am6}. The DC power balance is captured in \eqref{eqn:am15}. Grid-forming (GFM) inverters with droop control enable direct voltage and frequency regulation. In this architecture, the frequency-active power ($\omega$-P) droop mechanism produces angle setpoints \eqref{eqn:sm3}, \eqref{eqn:am1.5}, and frequency setpoints \eqref{eqn:am1}, while the voltage droop mechanism determines voltage setpoints as described in \eqref{eqn:am2}. These setpoints are generated using filtered power signals \eqref{eqn:sm1}-\eqref{eqn:sm2}, obtained by sensing the actual values as specified in \eqref{eqn:am13}-\eqref{eqn:am14}. 
Table \ref{tab:tgfm} presents the GFM parameters used in the simulations, adapted from \cite{qoria2020current} with additional modifications. Unless specified otherwise, all parameters are expressed in per unit.
\begin{table}[ht!]
    \centering
  \setlength\tabcolsep{3.5pt}
  \setlength\extrarowheight{2pt}
  \caption{Grid-forming (GFM) inverter parameters}
    \begin{tabular}{cccccccc}
      \Xhline{2\arrayrulewidth} 
      $V_0$ & $\omega_0$ & $R_f$ & $L_f$ & $C_f$ & $K_P$ & $K_Q$ & $\omega_{b}$ \vspace{0.8ex}\\
      \hline
      1 & 1 & 0.0072 & 0.05 & 0.3 & 0.1\% & 0.01\% & 1 \\
      \hhline{|========|}
      $\omega_{pc}$ & $\omega_{qc}$ & $K^P_{VC}$ & $K^I_{VC}$ & $K^F_{VC}$ & $K^P_{CC}$ & $K^I_{CC}$ & $K^F_{CC}$ \vspace{0.8ex}\\ 
       \hline
       332.8 rad/s & 732.8 rad/s & 1 & 1.16 & 1 & 2.5 & 1.19 & 0 \\ 
      \Xhline{2\arrayrulewidth} 
    \end{tabular}
    \label{tab:tgfm}
\end{table}
\subsubsection{Nomenclature}
The variables and parameters used in the differential and algebraic equations are indexed below.\\
\texttt{Dynamic states (11):}\\
\begin{small}
$x_1\ :\ \tilde{p} \rightarrow$ Filtered active power\\
$x_2\ :\ \tilde{q} \rightarrow$ Filtered reactive power\\
$x_3\ :\ \theta \rightarrow$ Load angle\\
$x_4\ :\ \beta_d \rightarrow$ d-axis voltage controller state\\
$x_5\ :\ \beta_q \rightarrow$ q-axis voltage controller state\\
$x_6\ :\ \gamma_d \rightarrow$ d-axis current controller state\\
$x_7\ :\ \gamma_q \rightarrow$ q-axis current controller state\\
$x_8\ :\ v_{c,D} \rightarrow$ (Global) D-axis filter output voltage $v_c$\\
$x_9\ :\ v_{c,Q} \rightarrow$ (Global) Q-axis filter output voltage $v_c$\\
$x_{10}\ :\ i_{t,d} \rightarrow$ (Local)  d-axis IBR output current $i_t$\\
$x_{11}\ :\ i_{t,q} \rightarrow$ (Local) q-axis IBR output current $i_t$
\end{small}
\newline\newline
\texttt{Algebraic states (20):}\\
\begin{small}
$y_1\ :\ \omega \rightarrow$ Angular frequency\\
$y_2\ :\ \Delta \omega \rightarrow$ Angular frequency deviation from setpoint\\
$y_3\ :\ v^{*}_{c,d} \rightarrow$ d-axis voltage setpoint $v^{*}_c$\\
$y_4\ :\ v^{*}_{c,q} \rightarrow$ q-axis voltage setpoint $v^{*}_c$\\
$y_5\ :\ i_{t,d}^{*} \rightarrow$ d-axis current setpoint $i_t^{*}$\\
$y_6\ :\ i_{t,q}^{*} \rightarrow$ q-axis current setpoint $i_t^{*}$\\
$y_7\ :\ i_{g,d} \rightarrow$ d-axis filter output current $i_g$\\
$y_8\ :\ i_{g,q} \rightarrow$ q-axis filter output current $i_g$\\
$y_9\ :\ i_{g,D} \rightarrow$ D-axis filter output current $i_g$\\
$y_{10}\ :\ i_{g,Q} \rightarrow$ Q-axis filter output current $i_g$\\
$y_{11}\ :\ i_{t,D} \rightarrow$ D-axis IBR output current $i_t$\\
$y_{12}\ :\ i_{t,Q} \rightarrow$ Q-axis IBR output current $i_t$\\
$y_{13}\ :\ v_{c,d} \rightarrow$ d-axis filter output voltage $v_c$\\
$y_{14}\ :\ v_{c,q} \rightarrow$ q-axis filter output voltage $v_c$\\
$y_{15}\ :\ p \rightarrow$ Active power output from filter\\
$y_{16}\ :\ q \rightarrow$ Reactive power output from filter\\
$y_{17}\ :\ v_{t,d} \rightarrow$ d-axis IBR output voltage $v_t$\\
$y_{18}\ :\ v_{t,q} \rightarrow$ q-axis IBR output voltage $v_t$\\
$y_{19}\ :\ u_{dc} \rightarrow$ DC-link voltage\\
$y_{20}\ :\ i_{dc} \rightarrow$ DC-link current
\end{small}
\newline\newline
\texttt{Parameters (17):}\\
\begin{small}
$p_1\ :\ p^{*} \rightarrow$ Active power setpoint \\
$p_2\ :\ q^{*} \rightarrow$ Reactive power setpoint \\
$p_3\ :\ \omega_0 \rightarrow$ Angular frequency setpoint of IBR \\
$p_4\ :\ V_0 \rightarrow$ Voltage setpoint of IBR \\
$p_5\ :\ K_P \rightarrow$ Active ($\omega$-P) droop coefficient \\
$p_6\ :\ K_Q \rightarrow$ Reactive (V-Q) droop coefficient \\
$p_7\ :\ \omega_{pc} \rightarrow$  Active power filter 3dB cut-off frequency \\
$p_8\ :\ \omega_{qc} \rightarrow$  Reactive power filter 3dB cut-off frequency \\
$p_9\ :\ \omega_b \rightarrow$ Base angular frequency \\
$p_{10}\ :\ K^P_{VC} \rightarrow$  Proportional gain of voltage controller \\
$p_{11}\ :\ K^I_{VC} \rightarrow$  Integral gain of voltage controller \\
$p_{12}\ :\ K^F_{CC} \rightarrow$  Feed-forward gain of voltage controller \\
$p_{13}\ :\ K^P_{CC} \rightarrow$  Proportional gain of current controller \\
$p_{14}\ :\ K^I_{CC} \rightarrow$  Integral gain of current controller \\
$p_{15}\ :\ K^F_{VC} \rightarrow$  Feed-forward gain of current controller \\
$p_{16}\ :\ C_f \rightarrow$ Filter capacitance \\
$p_{17}\ :\ L_f \rightarrow$ Filter inductance 
\end{small}

\subsubsection{System Description}
The differential-algebraic equations (DAEs) outlined below describe the system's behaviour.\\ \noindent
\texttt{Differential equations:}
\begin{flalign}
       \dot{\tilde{p}} &\ =\ -\omega_{pc} \tilde{p} + p \omega_{pc} \label{eqn:sm1} &&
        \\
        \dot{\tilde{q}} &\ =\ -\omega_{qc} \tilde{q} + q \omega_{qc} \label{eqn:sm2}
        \\
        \dot{\theta} &\ =\ \omega_b \Delta\omega  \label{eqn:sm3}
        \\
        \dot{\beta_d} &\ =\ v^{*}_{c,d} - v_{c,d} \label{eqn:sm4}
        \\
        \dot{\beta_q} &\ =\ v^{*}_{c,q} - v_{c,q} \label{eqn:sm5}
        \\
         \dot{\gamma_d} &\ =\ i^{*}_{t,d} - i_{t,d} \label{eqn:sm6}
        \\
         \dot{\gamma_q} &\ =\ i^{*}_{t,q} - i_{t,q} \label{eqn:sm7}
         \\
        \dot{v_{c,D}} &\ =\ \omega v_{c,Q} + \frac{1}{C_f}(i_{t,d} - i_{g,d}) \label{eqn:sm8}
        \\
         \dot{v_{c,Q}} &\ =\ -\omega v_{c,D} + \frac{1}{C_f}(i_{t,q} - i_{g,q}) \label{eqn:sm9}
        \\
        \dot{i_{t,d}} &\ =\ \omega i_{t,q} + \frac{1}{L_f}(v_{t,d} - v_{c,d})
        \label{eqn:sm10}
        \\
        \dot{i_{t,q}} &\ =\ -\omega i_{t,d} + \frac{1}{L_f}(v_{t,q} - v_{c,q})
        \label{eqn:sm11}
\end{flalign}

\noindent
\texttt{Algebraic equations:}
\begin{align}
        &\ \omega - \omega_0 - \Delta \omega = 0 \label{eqn:am1}
        \\&\
        \Delta \omega - K_P(p^{*}-\tilde{p}) = 0 \label{eqn:am1.5}
        \\&\
        v^{*}_{c,d} - V_0 - K_Q(q^{*} - \Delta q) = 0 \label{eqn:am2}
        \\&\
        i^{*}_{t,d} \!-\! K^F_{VC}i_{g,d} \!-\! K^P_{VC}(v^{*}_{c,d} \!-\! v_{c,d}) \!-\! K^I_{VC}\beta_d \!+\! v_{c,q} \omega C_f = 0 \label{eqn:am3}
\end{align}
\begin{align}
        &\
        i^{*}_{t,q}  \!-\! K^F_{VC}i_{g,q}  \!-\! K^P_{VC}(v^{*}_{c,q}  \!-\! v_{c,q})  \!-\! K^I_{VC}\beta_q  \!-\! v_{c,d} \omega C_f = 0 \label{eqn:am4}
        \\&\ v_{t,d}  \!-\! K^F_{CC}v_{c,d}  \!-\! K^P_{CC}(i^{*}_{t,d}  \!-\! i_{t,d})  \!-\! K^I_{CC}\gamma_d  \!+\! i_{t,q} \omega L_f = 0 \label{eqn:am5}
        \\&\
        v_{t,q}  \!-\! K^F_{CC}v_{c,q}  \!-\! K^P_{CC}(i^{*}_{t,q}  \!-\! i_{t,q})  \!-\! K^I_{CC}\gamma_q  \!-\! i_{t,d} \omega L_f = 0 \label{eqn:am6}
        \\&\ 
        i_{g,d} - i_{g,D}\cos(\theta) - i_{g,Q}\sin(\theta) = 0 \label{eqn:am7}
        \\&\
        i_{g,q} + i_{g,D}\sin(\theta) - i_{g,Q}\cos(\theta) = 0 \label{eqn:am8}
        \\&\ 
        i_{t,d} - i_{t,D}\cos(\theta) - i_{t,Q}\sin(\theta) = 0 \label{eqn:am9}
        \\&\
        i_{t,q} + i_{t,D}\sin(\theta) - i_{t,Q}\cos(\theta) = 0 \label{eqn:am10}
        \\&\ 
        v_{c,d} - v_{c,D}\cos(\theta) - v_{c,Q}\sin(\theta) = 0 \label{eqn:am11}
        \\&\
        v_{c,q} + v_{c,D}\sin(\theta) - v_{c,Q}\cos(\theta) = 0 \label{eqn:am12} 
        \\&\
        p - v_{c,d}i_{g,d} - v_{c,q}i_{g,q} = 0 \label{eqn:am13}
        \\&\
        q - v_{c,q}i_{g,d} + v_{c,d}i_{g,q} = 0 \label{eqn:am14}
        \\&\
        u_{dc}i_{dc} - v_{t,d}i_{t,d} - v_{t,q}i_{t,q} = 0 \label{eqn:am15}
\end{align}

\end{document}